%% file: main.tex
\documentclass[%
 reprint,
superscriptaddress,
showpacs,preprintnumbers,
nofootinbib,
 amsmath,amssymb, 
 aps,
 prd,
 longbibliography,
]{revtex4-1}

\usepackage{cancel}
\usepackage{accents}
\usepackage{mciteplus,slashed}
\usepackage{amssymb,cancel,amsmath}
\usepackage{dcolumn}
\usepackage{bm}
\usepackage[caption=false]{subfig}
\usepackage{appendix}
\usepackage{physics}
\usepackage{feynmp-auto}
\usepackage[T1]{fontenc}	
\usepackage{csvsimple}
\usepackage{hyperref}
\usepackage[section]{placeins}
\usepackage[capitalise]{cleveref}
\usepackage{booktabs}
\usepackage{graphicx}
\usepackage{mathrsfs}
\usepackage{enumitem}
\usepackage{babel,blindtext}

\graphicspath{{Figures/}}

\usepackage[dvipsnames]{xcolor}
\usepackage[normalem]{ulem}

\input{Commands/ryansCommands.tex}

\input{Commands/flags.tex}
\begin{document}

\title{
Millicharged Cosmic Rays and Low Recoil Detectors}

\author{Roni Harnik}
\email{roni@fnal.gov}
\affiliation{Theoretical Physics Department, Fermilab, Batavia, IL 60510,USA}

\author{Ryan Plestid}
\email{rpl225@uky.edu}
\affiliation{Department of Physics and Astronomy, University of Kentucky,  Lexington, KY 40506, USA}
\affiliation{Theoretical Physics Department, Fermilab, Batavia, IL 60510,USA}

\author{Maxim Pospelov}
\email{pospelov@umn.edu}
\affiliation{School of Physics and Astronomy, University of Minnesota, Minneapolis, MN 55455, USA}
\affiliation{William I. Fine Theoretical Physics Institute, School of Physics and Astronomy, University of Minnesota, Minneapolis, MN 55455, USA}
\author{Harikrishnan Ramani}
\email{hramani@stanford.edu}
\affiliation{Berkeley Center for Theoretical Physics, Department of Physics,
University of California, Berkeley, CA 94720}
\affiliation{
Theoretical Physics Group, Lawrence Berkeley National Laboratory, Berkeley, CA 94720}
\affiliation{Stanford Institute for Theoretical Physics,
Stanford University, Stanford, CA 94305, USA}

\date{\today}

\preprint{FERMILAB-PUB-20-523-T}
\begin{abstract}
    We consider the production of a ``fast flux'' of  hypothetical millicharged particles (mCPs) in the interstellar medium (ISM). We consider two possible sources induced by cosmic rays:  (a)  $pp\rightarrow$(meson)$\rightarrow$(mCP) which adds to atmospheric production of mCPs, and (b) cosmic-ray upscattering on a millicharged component of dark matter. 
    We notice that the galactic magnetic fields retain mCPs for a long time, 
    leading to an enhancement of the fast flux by many orders of magnitude. In both scenarios, we calculate the expected signal for direct 
    dark matter detection aimed at electron recoil. 
    We observe that in Scenario (a)  neutrino detectors (ArgoNeuT and Super-Kamiokande) still provide superior sensitivity compared to dark matter detectors (XENON1T). However, in scenarios with a boosted dark matter component, the dark matter detectors perform better, given the enhancement of the upscattered flux at low velocities. Given the uncertainties, both in the flux generation model and in the actual atomic physics leading to electron recoil, it is still possible that the XENON1T-reported excess may come from a fast mCP flux, which will be decisively tested with future experiments. 
\end{abstract}

 \maketitle 
 
 \section{Introduction \label{intro}}

Cosmic rays exhibit a rich phenomenology in the Earth's local neighborhood. Highly boosted charged particles strike protons in the interstellar medium and in the upper atmosphere, to produce showers of more numerous (and less energetic) charged particles. 
Any charged particle which is sufficiently light and long-lived will be present in cosmic ray showers, including the positron, the muon and the pion which were first discovered there \cite{Anderson:1932zz,Neddermeyer:1937md,Lattes:1947mw}.
In addition, charged cosmic rays interact with local magnetic fields and undergo a random walk, described well by a diffusion process \cite{Strong:2007nh}, which leads to an enhancement in their abundance near the galactic disk. As a result, cosmic rays, particularly slow moving particles, are retained in the vicinity of the galactic disk. 
Like any charged particles, incoming CRs will interact with matter through Rutherford scattering, $\dd \sigma/\dd T_e \propto 1/T_e^2$, with soft collisions being most abundant. 
Upon reaching the dense environment of the atmosphere, the Earth itself, or a detector, CRs will loose energy rapidly. This allows for their detection with tracking and calorimetry, but also allows one to protect detectors from CR background with shielding and underground laboratories.

Millicharged particles (mCPs) are a simple extension of the Standard Model which introduce just that, a new particle with a very small charge. They can arise in simple frameworks in which a new gauge boson kinetically mixes with the photon~\cite{Holdom:1985ag}. mCPs in the MeV to several GeV range have been identified as interesting targets for experimental analyses, with limits set by SLAC's milliQ~\cite{Prinz:1998ua}, milliQan  
\cite{Ball:2016zrp,Ball:2020dnx}, LSND \cite{Magill:2018tbb,Auerbach:2001wg}, Fermilab's miniBoone~\cite{Magill:2018tbb,Aguilar-Arevalo:2018wea} and ArgoNeuT~\cite{Harnik:2019zee,Acciarri:2019jly}, and proposed dedicated experiments~\cite{Kelly:2018brz,Liang:2019zkb,Choi:2020mbk,Foroughi-Abari:2020qar}. Importantly, for this paper, mCPs exhibit all of the same phenomena as cosmic rays, only to a lesser degree due to their small charge.

In addition to serving as a simple extension of the SM, mCPs can also have important astrophysical and cosmological consequences. mCPs have been invoked to explain several experimental excesses/anomalies ~\cite{Foot:2007cq,gies2006polarized, Wallemacq:2014sta,farzan2019dark,Khan:2020vaf,Farzan:2020dds}. Earlier work has considered connections of mCPs to dark matter and its observable consequences \cite{Dimopoulos:1989hk,Dubovsky:2001tr,Dubovsky:2003yn,Lepidi:2008hb,Cheung:2007ut,Berezhiani:2008gi}. More recently, there has been a resurgence of interest in mCPs, and specifically millicharged dark matter (mDM) \cite{Liu:2019knx,Berlin:2018sjs,Munoz:2018pzp,Barkana:2018qrx,Slatyer:2018aqg,Kovetz:2018zan,Creque-Sarbinowski:2019mcm,Munoz:2018jwq,Stebbins:2019xjr, Kachelriess:2020ams}, since the EDGES collaboration reported an anomaly in their 21 cm data \cite{Bowman:2018yin}. The proposal is that the enhancement of mCP-electron scattering at low velocities allow for a late thermal re-coupling of baryons and dark matter, lowering the kinetic temperature of atoms, and thereby enhancing the 21 cm absorption feature. This would be in line with stronger-than expected absorption claimed to be seen by the EDGES collaboration. Because the universe is awash with electromagnetic fields, the ultimate fate of mDM is a non trivial question and the current literature is in a state of flux with no clear consensus on either the velocity or density distributions of mDM. Some authors have argued that mDM is evacuated from the disk entirely by supernova shocks \cite{Dimopoulos:1989hk} so as to be unobservable in terrestrial experiments \cite{Chuzhoy:2008zy,McDermott:2010pa}, while others have argued that the same supernova shocks, coupled with diffusive transport eventually lead to an easily observable high-velocity distribution of mDM although estimates vary by orders of magnitude \emph{c.f.} \cite{Hu:2016xas} v.s.\ \cite{Dunsky:2018mqs}. More recent work has suggested that there are a variety of unaccounted for collective damping effects that could severely limit the speed of mDM reaching a present-day terrestrial detector \cite{Li:2020wyl}.  What is unambiguous, however, is that if mDM existed at the cosmic dawn and is responsible for the EDGES anomaly, then there should be a predictable (albeit model-dependent) density of mDM spread somewhere within roughly 20 kpc of our local galaxy.  

Motivated by the uncertainty in the predictions for a mDM density and velocity distribution, it is natural to ask if there is a direct-detection scheme that is relatively \emph{insensitive} to the details of mCP transport. We answer this question in the affirmative by considering two qualitatively different scenarios: 
\begin{enumerate}[label=(\alph*)]
    \item The mCPs \emph{do not} constitute a component of the dark matter, but are readily produced via collisions of cosmic rays with the interstellar medium (ISM) in addition to local cosmic-rays' collisions with the upper atmosphere (as discussed in \cite{Plestid:2020kdm}). 
    \item The mCPs \emph{do} form a sub-component of the dark matter in the ISM and are boosted via cosmic ray collisions. We refer to this scenario as ``Rutherford-boosted'' dark matter (RBDM). 
\end{enumerate}
\begin{figure}\raggedright
    \includegraphics[width=0.85\linewidth]{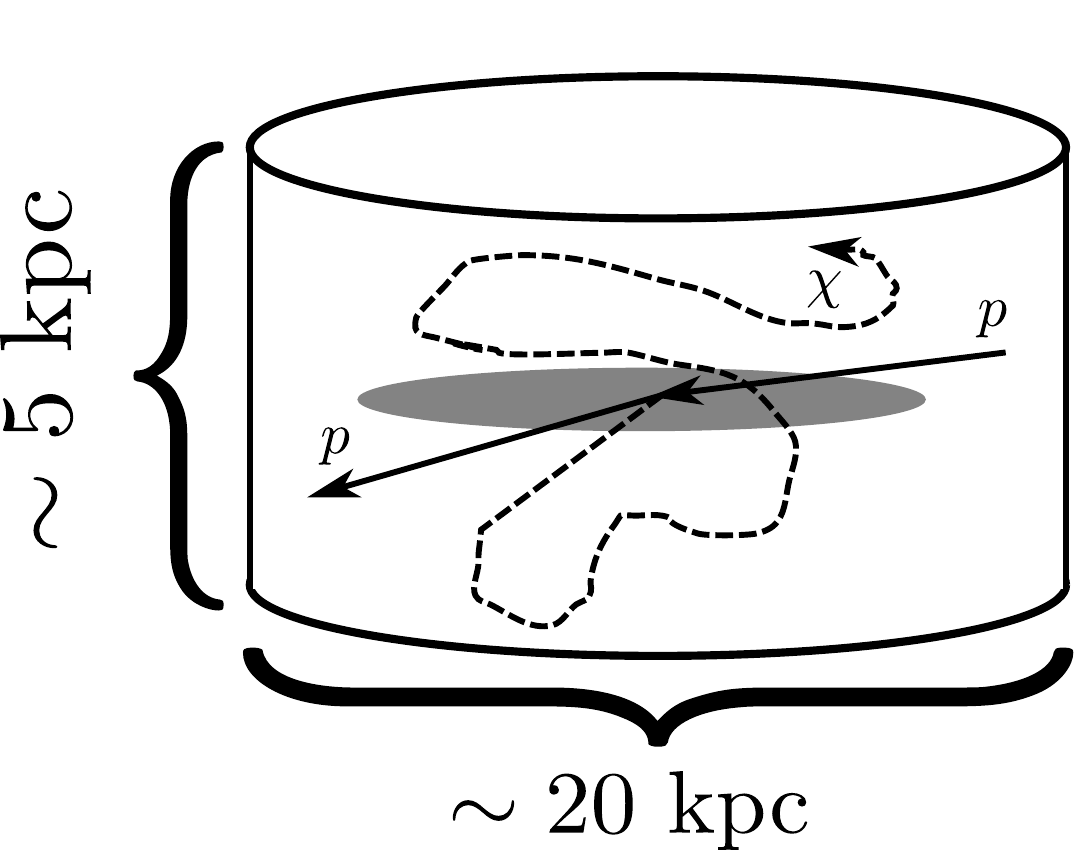}
    \caption{A cartoon depiction of the leaky-box/disk-diffusion model \cite{Ptuskin:2006xz}. A cosmic ray collides with a proton in the ISM producing an mCP pair via an intermediate meson. This results in a fast-moving mCP that diffuses slowly about the leaky-box. The long-retention time due to magnetic diffusion results in a greatly enhanced flux of mCPs sourced from the ISM \cite{Aloisio:2017ooo,Strong:2007nh}. \label{fig:leaky-box}}
\end{figure}
Scenario (a) is entirely independent of mDM modelling assumptions and can be used to set limits on mCPs in the same way a fixed-target or collider experiment could. Scenario (b) assumes an ambient population of mDM, but is relatively insensitive to the details of the number density and velocity distributions. Even if mDM is slow, cosmic-ray collisions will generate a RBDM component that is fast. Even if the mDM is evacuated from the disk, provided it lies within the cosmic-ray diffusion zone, cosmic rays will find it. Finally, even if collective damping effects are important for \emph{coherent} mDM motion, the RBDM component will be \emph{incoherent} having been generated by random and isotropic collisions throughout the volume of the cosmic ray diffusion zone. 

In both scenarios our results rely crucially on the diffusive motion of mCPs (see \cref{fig:leaky-box}) which leads to a residency time in our local galaxy that is four orders of magnitude larger than would be expected if the mCPs were to exhibit ballistic motion. This can result in the ISM component of the flux being \emph{larger} by an order of magnitude then the locally generated flux from the upper atmosphere despite the ISM being a ``thin target'' and the atmosphere being a ``thick target''. 

Rates of mCP detection naturally benefit from low-thresholds, which both increase the detection cross section and allow an experiment to access low-velocity components of an mCP flux. The XENON1T collaboration has demonstrated the ability to conduct low-background tonne-scale electron recoil experiments with a threshold in the 100-eV ($S_2$) \cite{Aprile:2019xxb} to few keV scale ($S_1$-$S_2$) \cite{Aprile:2020tmw}.  As we show below, such low thresholds are especially advantageous for Scenario (b). A natural question, then, is whether or not XENON1T's unique low-threshold capabilities are able to overcome the large discrepancies in fiducial volume between itself and Super-Kamiokande (1 t vs 22.5 kt), which has recently been identified as  a resource for studying mCPs from cosmic rays \cite{Plestid:2020kdm}. 
 
Our main results demonstrate that in Scenario (a), single electron recoil in the Super-Kamiokande still provides superior sensitivity even compared to the most sensitive dark matter detectors, such as XENON1T. We update the sensitivity/exclusion curves by accounting for the additional channel of mCP production in collisions of CRs with the interstellar medium. In Scenario (b) we exploit the fact that the flux of RBDM has a significant infrared enhancement, and are able to derive competitive limits from the XENON1T experiments' electron recoil analysis. Given enough uncertainty in the current treatment of atomic physics, we find that the claimed few keV-scale excess of recoil electrons may still plausibly come from the fast flux of millicharged particles, but this subject requires a more detailed investigation. Departing from these two relatively conservative scenarios, we explore the sensitivity to a fast mCP flux of more exotic origin, such as e.g.\ dark sector decay to mCP, finding that in this case the sensitivity may extend to very small values for the millicharge. 

The rest of the paper is organized as follows: in  \cref{mag-ret} we introduce the well-known, but crude, ``leaky-box'' model of magnetic retention that allows the ISM to contribute fluxes comparable to those coming from the upper atmosphere. Then, in \cref{fluxes} we outline the calculation of cosmic-ray fluxes from 1) the upper atmosphere and 2) the ISM. We also discuss the resulting RBDM flux in the case where mCPs form a fractional population of the DM. Next, in \cref{exclusions} we describe how event rates at the XENON1T detector are calculated and present exclusions and projections for both Scenario (a) and Scenario (b), as well as a generalization to the fast mCP flux of unspecified origin. In \cref{attenuation}, attenuation due to the Earth overburden is discussed. This introduces novel $\epsilon$ dependent scaling in the case of Scenario (b). Finally in \cref{conclusions} we summarize our results and comment on the future impact of low recoil detectors for mCP and mDM searches.  

\section{Magnetic Retention \label{mag-ret} } 

It is well known that magnetic fields lead to a very large enhancement in the cosmic ray intensity \cite{Aloisio:2017ooo,Strong:2007nh}. The essential physics is that magnetic fields scatter particles in such a way that they undergo a random walk described by a diffusion coefficient, $D$,  that depends on the magnetic rigidity, $R=p/|q|$  of the particle at hand (more on this later). This statement is robust, and so the magnetic retention of mCPs can be readily obtained by re-scaling the well understood magnetic retention of protons. In what follows we do not attempt an elaborate treatment (as in e.g.\ GALPROP) relying instead on a simplified diffusion model. 

We will quote all of our results in terms of ``leaky-box'' or ``disk-diffusion'' model wherein a particle is envisioned to diffuse about a leaky-box roughly conceptualized as a cylinder of radius $\sim 20$ kpc and a height of $\sim 5-10$ kpc; a cartoon of is shown in \cref{fig:leaky-box}. This defines a length scale $\ell_\text{LB}$ which conveniently \emph{parameterizes} the enhancement of the mCP flux due to magnetic retention.  This parameterization can be extended to account for a more sophisticated treatment of the mCP diffusion as we discuss in \cref{Diffusion-Discussion}. 

The residency time within this container is given by $\tau_\text{LB}\sim 2\ell_\text{LB}^2/D$ where $D =  D_0 \beta (R/R_0)^\delta$ is the rigidity dependent diffusion constant and the factor of two comes from the conventional definition in a diffusion equation $(\partial_t + \tfrac12 D\nabla^2)f=0$. Here, $R_0$ is a reference rigidity and $\delta$ is an exponent that is somewhat model dependent; a popular choice that is both theoretically well motivated and empirically successful is based on Komolgorov scaling, $\delta=1/3$ \cite{Ptuskin:2006xz}. This can be compared to the residency time of free-passage $\tau_\text{FP}= \ell_\text{LB}/\beta$. Realistic values of $\ell_\text{LB}$ would range from $1-10$ kpc, while $D_0 \approx 3 \times 10^{28} \text{cm}^2\text{s}^{-1}$  and $R_0= |e| \times [ 1~\text{GeV}]$. 

The ratio of the leaky-box retention time $\tau_\text{LB}$, to time of free-passage (i.e.\ ballistic motion), $\tau_{FP}$ determine the rigidity-dependent magnetic-retention enhancement of the mCP flux
\begin{equation}\label{mag-ret-eq}
    \mathcal{R}_B= \frac{\tau_\text{LB}}{\tau_\text{FP}} 
    =  6 \times 10^4   \qty[\frac{\ell_\text{LB}}{10~\text{kpc}} ]  \qty[\frac{1~\text{GeV} \times  \epsilon}{P_\chi}]^{\delta}   ~.
\end{equation}
This demonstrates the dramatic (four orders of magnitude) enhancement of mCPs that are sourced in the ISM relative to the naive estimate neglecting diffusive motion. We use $\ell_\text{LB}$ to quantify the relative size of magnetic retention, and we (again) encourage the reader to interpret this a convenient parameterization of the enhanced mCP flux relative to the naive ``free-passage'' calculation rather than as a bonafide physical parameter.  A more precise definition of $\mathcal{R}_B$ accounting for the spatial distribution of targets, and the cosmic ray density is given in \cref{Diffusion-Discussion}.

Foreshadowing slightly, this enhancement easily compensates for the relative paucity of scattering targets allowing mCP production in the ISM to be competitive with local production in the upper atmosphere. For instance, taking a cross section of 40 mb=4 $\times 10^{-25}$ cm$^{2}$ and an ISM column density of $n_\text{ISM}^\perp=$(1 cm$^{-3}$)$\times$(1 kpc)$=3 \times 10^{21}$ cm$^{-2}$, we find a probability of inelastic scattering to be $\sim 10^{-3}$ easily enhanced to an $O(1)$ number after accounting for magnetic retention. 

Lastly, it is important to note that there are broadly two versions of mCPs in literature and magnetic retention works only for a subset of these models. In the first model, mCPs are charged directly under the SM U(1), hence the mCPs behave very similar to the SM charged particles at all length scales and the results of this section hold. Particles charged under a dark U(1) which kinetically mixes with the SM U(1) also carry effective charge when probed at energies $\omega \gg m_{A'}$, the mass of the dark U(1) photon. In this model, if the dark photon mass is much larger than the inverse galactic length scales, there is no magnetic retention. For this reason, the main results of our paper are applicable to the first model described above, but wherever relevant we also show results in the absence of magnetic retention to best describe limits on mCPs with a heavy dark photon mediator. 
\begin{figure}
    \includegraphics[width=\linewidth]{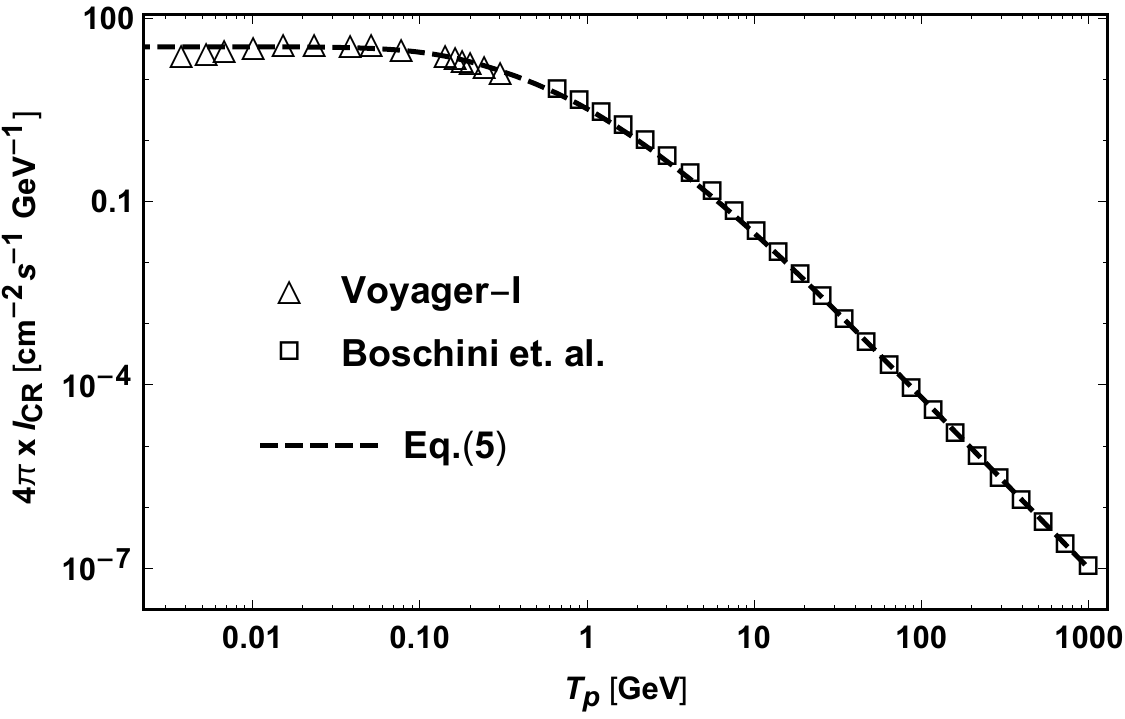}
    \caption{Comparison of \cref{simple-param} and the data from \cite{Cummings:2016pdr} (triangles) as well as the parameterization of \cite{Boschini:2017fxq} (squares). \label{Flux-Comparison}}
\end{figure}
\section{Fluxes of mCPs \label{fluxes}} 

Having discussed the retention of mCPs due to astrophysical magnetic fields, we now turn to their nascent production. As outlined in the introduction we consider two conceptually different scenarios and organize the following discussion accordingly.

\subsection{Scenario (a): cosmic-ray produced millicharges}

As has been recently demonstrated in \cite{Plestid:2020kdm}, cosmic rays can efficiently produce mCPs in the upper atmosphere. We consider the sensitivity of XENON1T to this production channel, but also consider a secondary flux component stemming from the ISM.  Much of the methodology used in this paper is outlined in detail in \cite{Plestid:2020kdm} and so we discuss only the most important details here and refer the interested reader to our earlier work.  

There are two main ingredients in the production of mCPs from cosmic ray collisions: 1) the production and resulting energy distribution of electromagnetically decaying mesons, and 2) the resultant spectral shape of the millicharged flux from the subsequent meson decays in flight. The former is much more challenging than the latter, but can be well approximated if one has accurate Feynman-$x$ distributions for the production of mesons in $pp$ collisions \cite{Plestid:2020kdm}. Once these are known it is a straightforward numerical exercise to boost the spectra into the lab frame weighted by the meson energy spectra obtained in step~1.

In the upper atmosphere a conservative estimate of the meson yield can be obtained by considering only primary production, the probability of which is given by the ratio of the meson-production to total-inelastic cross section $\sigma_\mathfrak{m}/\sigma_\text{inel}$. In the ISM things are slightly different and this factor should be replaced by $\sigma_\mathfrak{m} \nPerpISM$ where $\nPerpISM$ is the column density of the interstellar medium. This difference stems from the fact that the atmosphere is a thick target ($\lambda_\text{Atm}\sim$ 1 km with $\lambda=[\sigma n]^{-1}$ the mean free path) whereas the ISM is a thin target [$\lambda_\text{ISM} \approx 8 $ Mpc]. For the column density of the interstellar medium we take
\begin{equation}
    \nPerpISM =(1 ~\text{cm}^{-3}) \times (1~\text{kpc})~.
\end{equation}
This corresponds to the average density of protons in the ISM multiplied by $1~\text{kpc}$, which we take as a representative distance through which a cosmic ray will transverse the ISM.

In \cite{Plestid:2020kdm} spectra were quoted as differential with respect to $\gamma$. A simple change of variables (including a Jacobian) converts these spectra to being differential with respect to $\beta_\chi\gamma_\chi$. The mCP intensity can then be found (assuming isotropy) by multiplying the rate of production, by the column density of targets and the enhancement factor coming from magnetic retention, and dividing by $4\pi$
\begin{equation}\label{ISM-flux}
     \mathcal{I}^{\text{ISM}}_\chi= \frac{1}{4\pi}   \dv{R_\text{prod}}{\beta_\chi\gamma_\chi} \times n_\text{ISM}^\perp \times  \mathcal{R}_B~~.
\end{equation}
We note that while we have derived this expression within the context of an overly-simplified leaky-box model, a slightly more systematic treatment can be found in \cref{Diffusion-Discussion}, where $n_\text{ISM}^\perp \times  \mathcal{R}_B$ can be identified as the second bracketed term in \cref{generic-steady-state}. 

$R_\text{prod}$ is the production rate per ISM target particle, and is given by 
\begin{widetext}
\begin{equation}
    \dv{R_\text{prod}}{\beta_\chi\gamma_\chi} = 2 \sum_{\mathfrak{m}} \text{BR}(\mathfrak{m}\rightarrow \chi \bar\chi)\int \dd \gamma_\mathfrak{m} P(\beta_\chi\gamma_\chi|\gamma_\mathfrak{m} )\int \dd  T_p \mathcal{I}_\text{CR} (T_p) \sigma_{\mathfrak{m}}(T_p) P(\gamma_\mathfrak{m}|T_p)~,
\end{equation}
\end{widetext}
where the factor of two accounts for the combined $\chi$ and $\bar{\chi}$ flux. The conditional probabilities are determined by $\dd \sigma_\mathfrak{m}/\dd x_F$ and the details of $\mathfrak{m}\rightarrow \chi \bar\chi$ decays respectively~\cite{Plestid:2020kdm, Harnik:2019zee}. The flux of mCPs coming from the upper atmosphere is defined similarly but without the factor of $n_\text{ISM}^\perp \times \mathcal{R}_B$ in \cref{ISM-flux} and with $\sigma_{\mathfrak{m}}$ being divided by $\sigma_\text{in}$, the total inelastic $pp$ cross section. Importantly the flux from the upper atmosphere scales as $\epsilon^2$ since branching ratios involve two powers of $\epsilon$, whereas the flux from the ISM flux scales as $\epsilon^{2 +1/3}$ (for Komologorov scaling) and has a slightly different spectral shape due to the dependence of $\mathcal{R}_B$ on magnetic rigidity. 

\begin{figure}[t]
    \includegraphics[width=\linewidth]{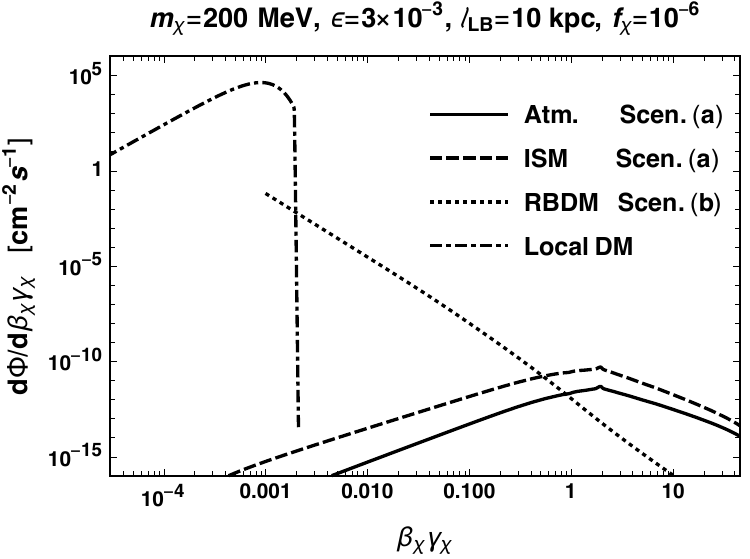}
    \caption{A comparison of the flux of mCPs arriving at earth produced from 1) Scenario (a): cosmic ray interactions in the upper atmosphere $\dd \Phi= 2\pi \mathcal{I}^\text{Atm}_\chi$,  2) Scenario (a): cosmic ray interactions in the ISM including magnetic retention (with $\ell_\text{LB}=10$ kpc) $\dd \Phi= 2\pi \mathcal{I}^\text{ISM}_\chi$,  3) Scenario (b): from RBDM with $f_\chi=10^{-6}$, $\dd \Phi= 2\pi \mathcal{I}^\text{RB}_\chi$, and 4) Virial dark matter: from the local dark matter density with a millicharged fraction of $f_\chi=10^{-6}$, $\dd \Phi = f_\chi n_\text{DM} v f(v)$ where $\int \dd v f(v) =1$. Benchmark values of $m_\chi=200$ MeV, and $\epsilon=3\times 10^{-3}$ have been chosen corresponding to viable parameter space to explain the EDGES anomaly as depicted in Fig.\ 1 of 
    \cite{Liu:2019knx}.  \label{fig:all-fluxes}}
\end{figure}

In Figure~\ref{fig:all-fluxes} we show the flux of mCPs in this scenario for $\epsilon=3\times 10^{-3}$, and $m_\chi=200$ MeV. The fluxes for production in the upper atmosphere and in the ISM (including magnetic retention) are shown. Both fluxes peak at a few GeV, but as advertised, the ISM component is about an order of magnitude larger and slightly softer.

Finally, we would like to emphasize that in the upper atmosphere we approximated $pA$ collisions as a sum of incoherent $pp$ collisions, and we neglected secondary and tertiary meson production, thus our approach is conservative. In the ISM, however, collisions genuinely are $pp$ collisions, and since the column density is much smaller than the inverse cross section, secondary production is negligible. Therefore our treatment of the ISM production is robust, with the only (but important and sizeable) uncertainties stemming from our treatment of magnetic retention.

\subsection{Scenario (b): Rutherford boosted dark matter}

Having established the ISM as a competitive source of mCPs from $pp$ collisions, it is interesting to posit that some fraction of DM is millicharged (mDM) such that, with no additional assumptions,  cosmic rays would boost this component of the DM via Rutherford scattering. Rutherford scattering is enhanced by a $1/\beta_p^2$ factor such that scattering via low-energy cosmic rays dominate the RBDM spectrum. It is therefore imperative that we use reliable data on the cosmic ray spectrum in the ISM at low (i.e. 10 MeV scale) energies. The CR flux in the ISM is known to differ substantially from measured ``top of atmosphere'' fluxes at low energies. Fortunately Voyager 1 has both measured the cosmic ray spectrum beyond the heliosphere \cite{Cummings:2016pdr} and provided the data in the energy domain we require to estimate the flux of RBDM. We use a simple analytic interpolation of their data for the cosmic ray spectrum in the ISM given by
\begin{equation}
\label{simple-param}
    \mathcal{I}_\text{Voy} = \frac{1}{4\pi} \times \qty[\frac{25~ \text{cm}^{-2}\text{s}^{-1}}{\sqrt{x^4+x^2+0.02}}] \qty[\frac{1}{\sqrt{x^{1.6}+8^{1.6}}}] ~,
\end{equation}
where $x=T_p/[1~\text{GeV}]$. This matches on to more sophisticated parameterizations at higher energies \cite{Boschini:2017fxq}, and reproduces the data reasonably well at lower energies as can be seen in \cref{Flux-Comparison}. Since we anticipate that the size of the magnetic retention introduces a much greater uncertainty in the flux, \cref{simple-param} is adequate for our purposes. 

The rate of RBDM production per mDM target is given by 
\begin{equation}
    \label{ruth-DM-flux}
  \dv{R_\text{prod}}{\beta_\chi\gamma_\chi} = \int_{T_\text{min}}^\infty \dd T_p ~ 
  4\pi \mathcal{I}_\text{Voy}(T_p)  \dv{ \sigma}{\beta_\chi\gamma_\chi }~. 
\end{equation}  
Here, $T_\text{min}$ is a function of $T_\chi$ that is found by demanding the incident proton have enough energy to kick an mCP to a kinetic energy of $T_\chi$. By solving for the maximum four-momentum transfer possible one can find
\begin{equation}\label{tmin-2}            
  T_\text{min}= \sqrt{\frac{
  \left(2 m_{\chi}+T_{\chi }\right) 
  \left(2 m_p^2+m_{\chi } T_{\chi }\right)}{4m_{\chi }}}
  +\tfrac12 T_{\chi }-m_p~.            
\end{equation}
The flux of RBDM is found, in direct analogy with \cref{ISM-flux}, by multiplying by the number density of mDM, and the magnetic enhancement factor 
\begin{equation}\begin{split}
    \dv{\mathcal{I}_\chi^\text{RB}}{\beta_\chi\gamma_\chi} &=\frac{1}{4\pi} \dv{R_\text{prod}}{\beta_\chi\gamma_\chi}  \times n_\text{mDM}^\perp \times \mathcal{R}_B
\end{split}\end{equation}
where we have parameterized the mDM column density as 
\begin{equation}
    n_\text{mDM}^\perp = (1~\text{cm}^{-3})\times (10~\text{kpc}) \times f_\chi \times (0.3 ~\text{GeV}/m_\chi)~.
\end{equation}
We have taken a longer column length for DM than for the ISM contribution (1 kpc vs 10 kpc) because while the local DM density is roughly equal to the local ISM density, DM is spread over a larger volume. In this paper we define the mDM fraction $f_\chi$ \emph{globally} with respect to the Milky Way
\begin{equation}
    f_\chi = \frac{ \int_\text{MW}\dd^3x~ \rho_\text{mDM}(x)}{\int_\text{MW}\dd^3x~ \rho_\text{DM}(x)}~.
\end{equation}
As discussed in the introduction, mDM transport can lead to over- or under-densities of mDM in our local vicinity. We can imagine two extreme cases: One could imagine that mDM falls back onto the disk \cite{Dunsky:2018mqs}, and has a large enough transport cross section with baryons that the mDM matter profile closely tracks the baryonic profile in the disk. Alternatively, if mDM is ejected it could be present in the cosmic-ray diffusion volume, but be well separated from the disk (for instance $\sim 5$ kpc out of the plane). We estimate that even in these extreme scenarios the re-distribution of the mDM density only leads to an $O(10)$ enhancement of the RBDM flux arriving at Earth (see \cref{Diffusion-Discussion}).

\begin{figure}[t]
    \includegraphics[width=\linewidth]{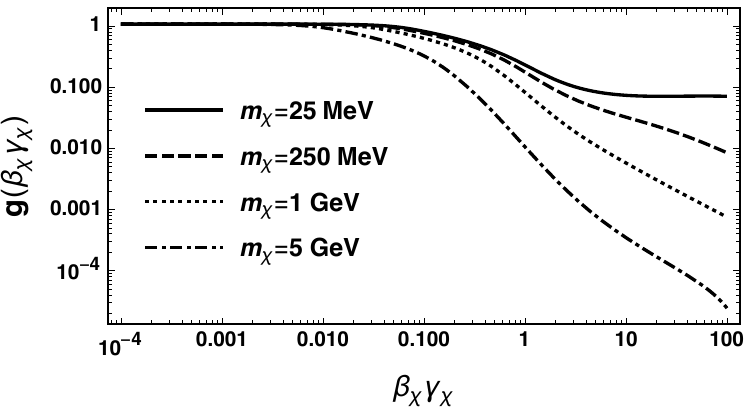}
    \caption{Behavior of $g(\beta_\chi\gamma_\chi,m_\chi)$ [see \cref{g-def}] for $m_\chi=$ 25 MeV, 250 MeV, 1 GeV, and 5 GeV. \label{Flux-Func}}
\end{figure}

We can transform from  $T_\chi$ to $\beta_\chi \gamma_\chi$ by including the Jacobian $\dd T_\chi/ \dd \beta_\chi\gamma_\chi$ and multiplying through by the magnetic retention factor $\mathcal{R}_B$
\begin{equation}\begin{split}
    \mathcal{I}_\chi^\text{RB} =\frac{1}{4\pi}&\mathcal{R}_B(\beta_\chi\gamma_\chi) \times  \dv{R_\text{prod}}{T_\chi} \times \dv{T_\chi}{\beta_\chi \gamma_\chi} \\
    &\times (10~\text{kpc}/\text{cm}^{3}) \times f_\chi \times \frac{0.3~\text{GeV} }{m_\chi} ~,
\end{split}\end{equation}
where $T_\chi= m_\chi(\gamma_\chi-1) = m_\chi (\sqrt{(\beta_\chi\gamma_\chi)^2 +1} -1)$. Putting everything together, and allowing mDM to be some fraction, $f_\chi$, of the total DM density,  we estimate the intensity of RBDM arriving at earth  to be 
\vfill
\begin{widetext}
\begin{equation}\label{g-def}
  \mathcal{I}_\chi^\text{RB}\approx (3.8 \times 10^{-7}~\text{cm}^{-2}~\text{s}^{-1} )   \bigg[\frac{f_\chi}{10^{-2}}\bigg]\bigg[\frac{\epsilon}{10^{-3}}\bigg]^{2+\tfrac13}\bigg[\frac{\ell_\text{LB}}{10~\text{kpc}}\bigg] \bigg[\frac{300~\text{MeV}}{m_\chi}\bigg]^{3+\tfrac13} \bigg[\frac{0.1}{\beta_\chi\gamma_\chi}\bigg]^{3+\tfrac13} g(\beta_\chi\gamma_\chi, m_\chi)
\end{equation}
\end{widetext}
where $g(\beta_\chi \gamma_\chi, m_\chi)$ is a dimensionless function that depends on $m_\chi$ and satisfies $g(0,m_\chi)=1$; \cref{Flux-Func} shows its behavior as a function of $\beta_\chi \gamma_\chi$ for a few representative masses. 

As an illustration of the various components of the flux arriving at a terrestrial detector (without considering effects of attenuation due to e.g.\ overburden or solar winds) we show the differential flux in  \cref{fig:all-fluxes} for the benchmark point where one part per million of dark matter is millicharged $f_\chi=10^{-6}$, $\epsilon=3\times 10^{-3}$, and $m_\chi=200$ MeV (this corresponds to an EDGES preferred point of parameter space as presented in Fig.\ 1 of \cite{Liu:2019knx}). We compare this to a Maxwellian velocity distribution cut-off at $575$ km/s. RBDM adds a hard tail of fast millicahrged dark matter, as expacted.

\subsection{Exotic Scenarios for large fluxes of millicharges \label{large-fluxes}}

Even larger fluxes of mCPs can be motivated in more exotic scenarios than the fairly minimal 
Scenarios (a) and (b) considered thus far and we pause here to discuss some possibilities. The question of maximal flux of exotic particles (mCP or other variety) is quite general and deserves a separate discussion. Some exotic sources of fast moving DM include boosted DM \cite{Agashe:2014yua,Kong:2014mia}, solar reflection \cite{An:2017ojc} and SM decays \cite{McKeen:2020vpf}. For concreteness we consider here, scenarios with minimal dark sector content. The sources of exotic flux with $v>0.1$ we consider are: {\em i.\ } dark matter decay globally and in the galaxy; {\em ii.\ } dark matter decay or annihilation inside astronomical objects such as the Sun and solar system planets; {\em iii.\ } exotic radioactivity from the decay of normal atoms.

Scenario {\em i.\ } posits that throughout a history of the Universe there existed, or still exists, an unstable agent that produces pairs of mCPs, according to $X\to 2{\rm mCP}$  (see \cite{Cui:2017ytb} for more extended discussion of similar scenarios). If $X$ constitutes the dark matter today, 5\% or so of decays after the recombination until now is permissible. Notice that, for example, $X$ can be a subdominant component of the dark matter that has fully decayed to mCPs. Then the mCP's spectrum, which is presumably (semi)-relativistic initially, can be significantly softened by the subsequent expansion.
Therefore one can find a situation where only the DM experiments would be sensitive to this type of flux, while for neutrino experiments the flux would be below the energy threshold. Scenario {\em ii.\ } occurs when DM is effectively intercepted by large astronomical bodies, such as thee Sun, with subsequent annihilation or decay. In this case, the maximum flux will be limited by the "supply" of DM, that is the maximum number of particles intercepted by the Sun with the subsequent annihilation to mCPs. Scenario {\em iii.\ } draws on recent ideas in \cite{McKeen:2020zni,McKeen:2020vpf} where Hydrogen atoms undergo decays to exotic particles, which in turn can lead to the mCPs. This scenario (H-atom decay) is somewhat less constrained that the one based on proton decays, as constraints on H-decay are less stringent than in the proton case (see \cite{McKeen:2020zni,McKeen:2020vpf} and references therein). 

In each of these three cases we can estimate the maximum amount of mCP flux that can be created. Assuming a GeV-mass progenitors for the mCPs, we get: 
\begin{align}
    \Phi_i &\leq 10^4{\rm cm}^{-2}{\rm s}^{-1}\left( 
\frac{10\tau_U}{\tau_{X}}\right)\left(\frac{1\,\rm GeV}{m_{X}}\right)\\
\Phi_{ii} &\leq 10^1\, {\rm cm}^{-2}{\rm s}^{-1}\left(\frac{1\,\rm GeV}{m_{X}}\right)\\
\Phi_{iii}  &\leq 10^3\, {\rm cm}^{-2}{\rm s}^{-1}\left(\frac{10^{28}\rm s}{\tau_{\rm H }}\right)~.
\end{align}
In comparison, the total virial DM flux is $\Phi_{\rm vir}\simeq 10^7{\rm cm}^{-2}{\rm s}^{-1}\times ({\rm GeV}/m_{\rm DM})$. Thus we see that the non-minimal options for obtaining ``fast mCP flux" could be as high as $\sim 0.1\%$ of the main DM flux. 

In addition to these exotic, but relatively straightforward, scenarios one can imagine slightly more elaborate mechanisms for generating a fast flux of mCs, although each of these will require further scrutiny. For example, a potentially interesting effect is related to the formation of nucleus-mCP bound states for the negatively charged particles. This possibility refers to larger mass and larger $\epsilon$ part of the parameter space. Taking, for example, helium nucleus into consideration, one can estimate the binding energy to be $\sim (\mu/(2\,{\rm GeV})\times (\epsilon/0.05)^2\times 133\,{\rm eV}$, where $\mu$ is the reduced mass of the mCP-Helium nucleus pair. Binding energies of this size will lead to post-BBN, but pre-recombination formation of bound states. These bound states, if surviving until today, can be accelerated at astrophysical sites almost as efficiently as other nuclei. Moreover, upon entering Earth’s atmosphere, the mCPs in such bound states can be easily stripped by nuclear collisions, creating a flux of energetic mCPs. Interestingly,  even if the binding of mCPs to nuclei is mediated by a massive dark photon, such that magnetic retention of RBDM is inhibited (because electromagnetism is shielded at large distances), this ``nuclear delivery system’’ mechanism would still be viable as it relies \emph{exclusively} on the non-zero electric charge of the He-mCP atom. While this mechanism is certainly interesting, it could potentially be prone to many uncertainties, and chiefly among them the abundance of bound states in specific astrophysical settings, several billion years after the Big Bang. Therefore we do not pursue it further in this paper. 

At smaller values of $\epsilon$, mCP-nucleus bound states will not form. However, one could imagine mCP-mCP bound states to be a dominant DM subcomponent provided that there are additional dark sector dynamics that bind mCPs to one another. For example  a dark photon could bind two oppositely charged mCPs (possibly of two different species with different masses as considered in \cite{Kahlhoefer:2020rmg}). Binding energies could be sufficiently large that bound states would form in advance of recombination, and in this case bounds from the CMB \cite{Kovetz:2018zan} would need to be reinterpreted. A substantial fraction of DM could then be composed of neutral mCP-mCP bound states, which would later be ionized by CR scattering events leading to a larger \emph{effective} value of $f_\chi \gg 0.004$. This would contribute to (and thereby enhance) the RBDM flux. The viability of this scenario is highly model dependent, and since it would require further dark sector dynamics to facilitate mCP-mCP bound state formation we do not consider it further here. 

In addition to these exotic scenarios, a larger-than-estimated enhancement from magnetic retention is easily envisioned. For example if the diffusion constant were to rise more steeply at low velocities than we have modelled (e.g. $D \sim D \beta^{\eta} (R/\text{1 GeV})^\delta$ with $\eta\leq 1$ as suggested in e.g.\ \cite{Putze:2010fr,Ptuskin:2005ax,DiBernardo:2010is,Cholis:2011un}) then the RBDM flux would be even further enhanced in the infrared. A similar effect could be achieved by considering Kraichnan ($\delta=1/2$) rather than Komologorov scaling. Also, as discussed in \cref{Diffusion-Discussion}, the distribution of mDM itself can lead to a large magnetic retention. If the mDM is constrained to lie in the galactic disk rather than in an NFW-like profile then the overlap of mDM targets with cosmic rays is greater, and the local flux is enhanced because all of the RBDM sources are closer to Earth.   

It is also possible that low-energy cosmic rays have a much larger intensity in substantial regions of the Milky Way than the measurements of the Voyager-I mission would suggest. Indeed, authors have speculated in the past that the cosmic ray spectrum could be several orders of magnitude larger at low energies, and that such a spectrum could explain unexpectedly large concentrations of $H_3^+$ \cite{Indriolo:2009tf}. Since it is the low-energy part of the cosmic-ray proton spectrum that dominates the RBDM flux, if the low energy (sub GeV) cosmic ray intensity were to be this large \emph{anywhere} in the galaxy with a significant mDM population, then it lead to  substantial increase in the  RBDM flux arriving at Earth relative to the estimates presented in this paper.   

\section{Results \label{exclusions}}

\begin{figure}
    \includegraphics[width=\linewidth]{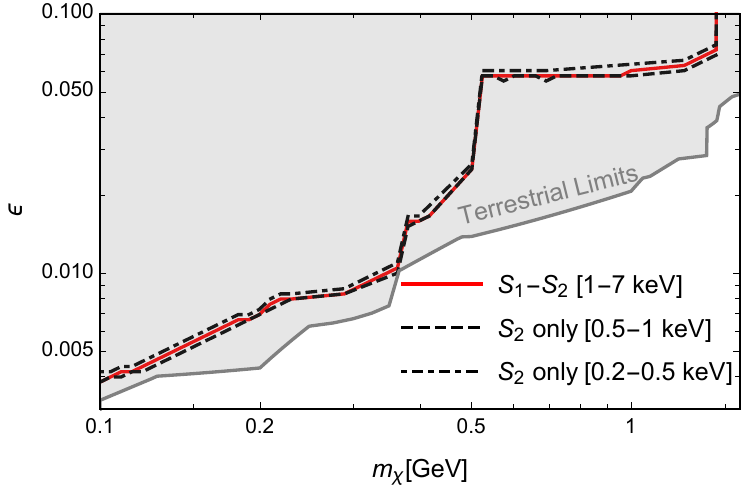}
    \caption{Scenario (a): Limits on mCPs arising from meson decays in the ISM from XENON1T data in the $S_2$ only bins \cite{Aprile:2019xxb} are shown in black. Also shown is the parameter space that explains the excess events in the 1-7 keV bins in $S_1$-$S_2$ data \cite{Aprile:2020tmw} in red. In gray we show a compilation of current terrestrial limits on MCPs \cite{Plestid:2020kdm} and also include the recent milliQan-demonstrator results \cite{Ball:2020dnx} .}
    \label{fig:crbinscomp} 
\end{figure}

\begin{figure*}
    \centering
   \subfloat[]{\includegraphics[width=0.495\linewidth]{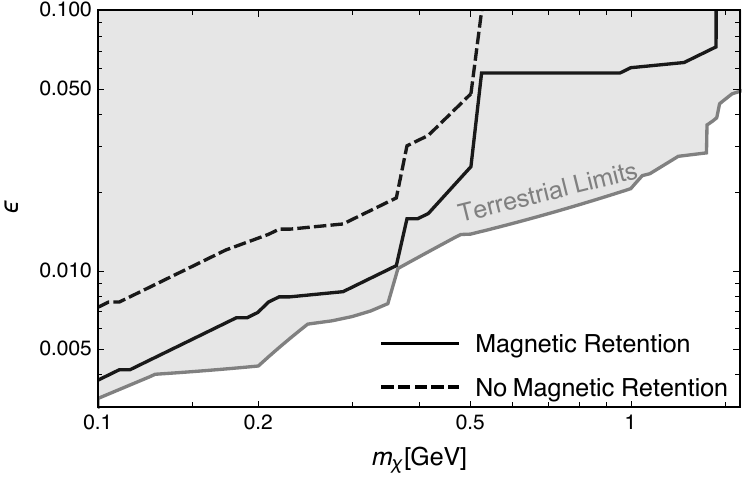}}
    ~ 
\subfloat[]{\includegraphics[width=0.495\linewidth]{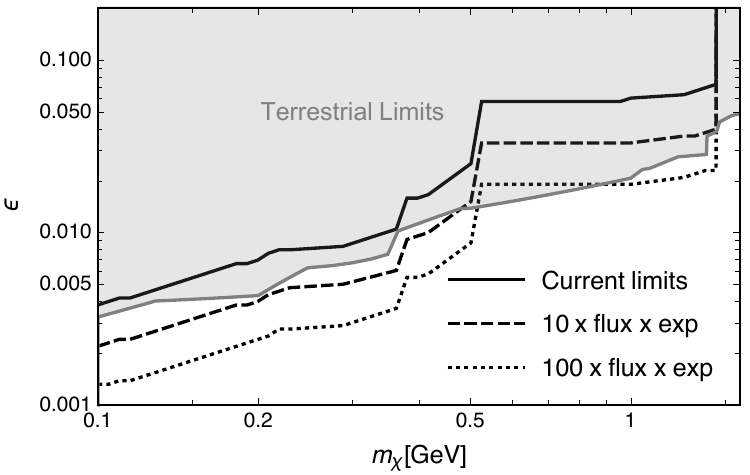}}
    \caption{Scenario (a): (\textbf{Left:}) Comparison of limits from the the $S_2$ only $\left(0.5 \textrm{keV}- 1 \textrm{keV} \right)$ bin with and without magnetic retention. (\textbf{Right:}) Projections for the $S_2$ only $\left(0.5 \textrm{keV}- 1 \textrm{keV} \right)$ bin for $10 \times$ and $100 \times$ the flux $\times$ experimental sensitivity.} 
    \label{crcomps}
\end{figure*}

Having established momentum dependent fluxes for both Scenario (a) and Scenario (b), we now turn to setting limits from the resultant rate in the electron ionization analyses in XENON1T as well as projections for the future. The differential cross section for ionization involves atomic physics form-factors which are described in \cref{atomic}. 
We estimate the rate of electron recoils in an energy bin as 
\begin{equation}
    R=\int_{E_{\rm min}}^{E_{\rm max}} \hspace{-3pt}\dd E_t \int_{\beta \gamma_{\rm min}}^\infty \hspace{-6pt} \dd \beta \gamma~2\pi\mathcal{I}_\chi \frac{d\sigma}{dE_t} \textrm{eff}(E_t)  ~,
\end{equation}
where $\textrm{eff}(E_t)$ is the electron efficiency as a function of energy transfer $E_t$ and the differential cross section $\frac{d\sigma}{dE_t}$ is provided in \cref{prescr}. The factor of $2\pi$ comes from the fact that we only consider zenith angles above the horizon. The minimum velocity is set by a combination of the experiment's electron recoil energy threshold and overburden attenuation. This latter effect is dependent on $\epsilon^2/m_\chi$ and introduces some non-trivial dependence of the rate on $\epsilon$ for Scenario (b) (RBDM) as shown in \cref{bgmin}. We therefore defer a discussion of the overburden until after presenting our results.

We consider two analyses from XENON1T in this work. First, we consider the $S_2$ only analysis from \cite{Aprile:2019xxb} with 22 tonne day exposure. Since the $S_2$ only analysis has poorly understood systematic uncertainties we only use it for limit setting purposes.
Following \cite{Bloch:2020uzh}, we take 22 events in the $\left(0.2-0.5\right)$ keV bin and 5 events in the $\left(0.5-1\right)$ keV bin as limiting rates (all parameter space leading to larger event counts is excluded). The efficiency $\textrm{eff}$ is assumed to be unity. The second analysis considered is the recently reported excess in the $S_1$-$S_2$ analysis from \cite{Aprile:2020tmw} with 0.65 tonne year exposure. The efficiency eff, is taken from Fig.2 of \cite{Aprile:2020tmw}. The collaboration observed 53 excess events in the $1-7$ keV bins with the excess peaked at the $\left(2-3\right)$ and $\left(3-4\right)$ keV bins.

\subsection{Scenario (a): Cosmic-ray produced mCPs}

As seen in \cref{fig:all-fluxes}, the flux of mCPs in Scenario (a) is peaked at (semi-)relativistic $\beta \gamma$. This allows almost the entire spectrum to penetrate the overburden of 3650 mwe, for the values of $\epsilon$ we consider here. As we will see in Sceanrio (b), at lower velocities the overburden plays an extremely important role in determining the experimental sensitivity. 

First, we compare limits from the different XENON1T analyses in \cref{fig:crbinscomp}. We show limits obtained from the two $S_2$ bins in black. The red curve corresponds to the 53 event excess in $1-7$ keV in the $S_1$-$S_2$ analysis. We find that for parameters capable of reproducing the 53 event excess in the $1-7$ keV bins, the expected signal in the 0.5-1 keV bin of the $S_2$ only analysis is roughly $6-7$ events (as against the 5 event limit). Important to our predictions is a treatment of the atomic physics governing the detector response that differs from commonly advocated approximations in the literature. We discuss the details of our approach in \cref{atomic} and point out that a plane-wave treatment with a Sommerfeld/Fermi function overestimates the event rate, and dramatically so at low recoil energies. 

In addition to (slightly) overpredicting the $S_2$ only event yield in the $0.5-1$ keV bin, the preferred parameter space of the $S_1$-$S_2$ analysis is firmly ruled out by existing constraints on mCPs. This includes the non observation of cosmic-ray produced mCPs from the upper atmosphere in Super-Kamiokande \cite{Plestid:2020kdm},  beam-induced mCPs in ArgoNeuT \cite{Acciarri:2019jly}, and recent results from the milliQan demonstrator 
\cite{Ball:2020dnx} (shown together in shaded gray in \cref{fig:crbinscomp}). 
\begin{figure*}[htpb]
    \centering
    \subfloat[]{\includegraphics[width=0.495\linewidth]{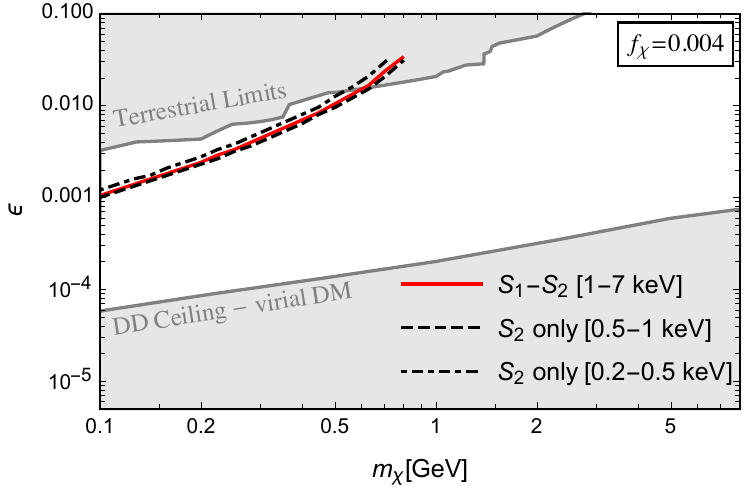}}
  ~ 
    \subfloat[]{\includegraphics[width=0.495\linewidth]{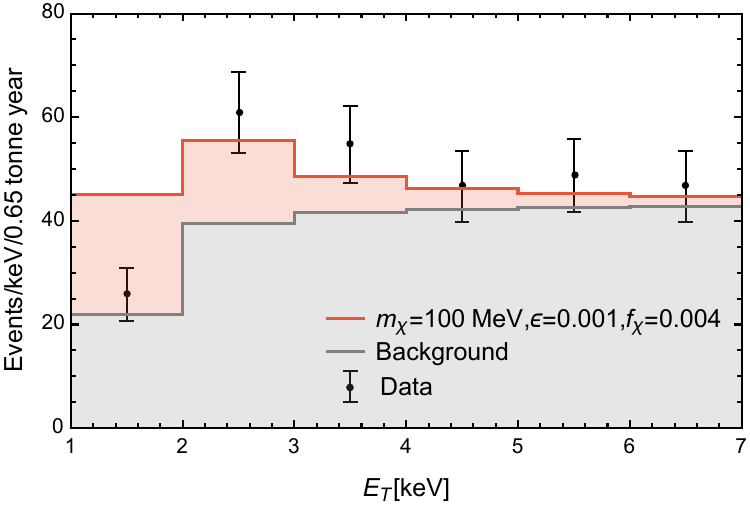}}
    \caption{Scenario (b). (\textbf{Left}): Limits on mCPs making up a fraction $f_\chi=0.004$ and accelerated by CR in the ISM, arising from XENON1T data in the $S_2$ only bins \cite{Aprile:2019xxb} are shown in black. Also shown is the parameter space that explains the excess events in the 1-7 keV bins in $S_1$-$S_2$ data \cite{Aprile:2020tmw} in red. In gray we show a compilation of current terrestrial limits on MCPs as well as DD limits and its ceiling \cite{Emken:2019tni}. (\textbf{Right}): Comparison of excess events for the parameter choice $m_\chi=100$ MeV, $\epsilon =10^{-3},$ and $f_\chi=0.004$, a sample point from the left panel with the data taken from \cite{Aprile:2020tmw}}. 
    \label{excessstuff}
\end{figure*}

\Cref{crcomps} (Left) compares the exclusions with and without magnetic retention for the $\left(0.5-1\right)$ keV bin ($S_2$ only). The modest gain in flux due to retention (relative to the component produced in the upper atmospheere) exhibited in \cref{fig:all-fluxes} translates to stronger exclusions. Limits on mCP models with a light, but not massless, dark photon are equivalent to the line without magnetic retention, because, although laboratory scattering events are unaffected, on large scales the dark photon mass screens astrophysical magnetic fields nullifying the effects of magnetic retention. 

Finally, in 
\cref{crcomps} (Right), we show projections for models that predict a higher flux, or for future analyses with higher experimental sensitivity. We find that if the experimental sensitivity can be improved by a factor of ten then new constraints can be set on mCPs independent of any assumptions about dark matter. We also note that this improved sensitivity could stem not only from improvements to the detector, but also if the flux were to be larger than estimated here because of e.g.\ a larger diffusion coefficient.

\subsection{Scenario (b): Rutherford boosted dark matter}
\begin{figure}
    \centering
    \includegraphics[width=\linewidth]{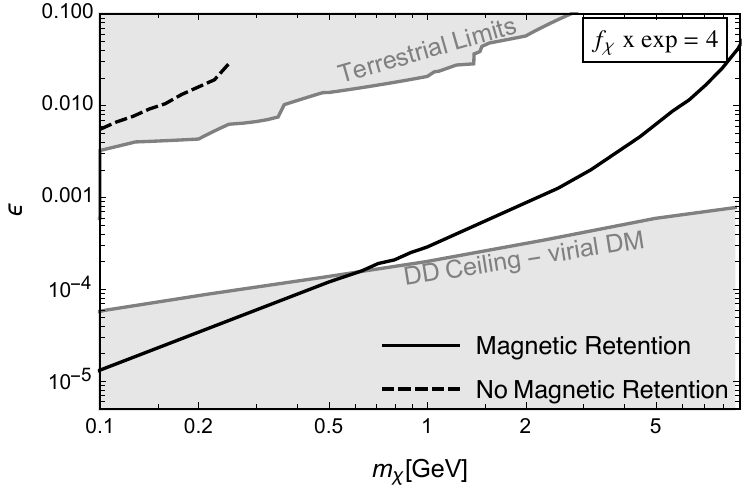}
    \caption{Scenario (b). Here we illustrate the dramatic consequences of magnetic retention by plotting the parameter space that yields an expected rate of 5 events in the $0.5-1$ keV bin of the $S_2$ only analysis. We show the parameter space accounting for magnetic retention (solid) relative to the parameter space that would be expected without magnetic retention (dashed). A RBDM flux that remains \emph{unenhanced} could occur if scattering is mediated by a light, but not massless, gauge boson such that astrophysicial magnetic fields are effectively screened.  We choose an unrealistically large concentration of mDM ($f_\chi \times\textrm{exp} =4$ with exp denoting experimental sensitivity) so that the unenhanced parameter space is visible to the reader. } 
     \label{fig:rbdmcomp-1}
\end{figure}

From \cref{fig:all-fluxes} we see that the RBDM flux is highly peaked at low velocities (scaling as $\beta_\chi^{-4.33}$) and there are two immediate consequences of this observation. First, unlike in Scenario (a), the sensitivity of XENON1T will greatly exceed that of Super-Kamiokande because of the huge gain in flux offered by the lower kinematical threshold on $\beta_\chi \gamma_\chi$. Second, the effect of overburden attenuation is more important here than in Scenario (a). At low velocities the incident mCPs have less energy (being more easily stopped in rock) and interact more strongly with matter (due to the $1/\beta^2$ scaling of the Rutherford cross section). 

All of our constraints depend on the fraction of DM that carries a millicharge, $f_\chi$. Constraints from the CMB power spectrum require that mDM compose only a small (sub-percent) fraction of DM \cite{Kovetz:2018zan}. A convenient benchmark parameter choice that is ``safe'' from the perspective of CMB constraints is $f_\chi=0.004$ and we adopt this as a benchmark value of $f_\chi$ throughout this section. This same parameter choice also ensures that the mDM we consider is not ruled out by constraints on self-interactions in galaxy cluster collisions \cite{robertson2016does}.  

We start by comparing the different XENON1T ionization analyses in \cref{excessstuff} (Left) for $f_\chi=0.004$. Much like in Scenario (a), we find that the  parameter space capable of producing a 53 event excess in the $S_1$-$S_2$  analysis (plotted in red) results in a slight overpopulation of the $0.5-1$ keV $S_2$ only bin (relative to the 5 events) and a slight under-population of the $0.2-0.5$ keV bin (relative to 22 events). Thus, the $S_2$ only and $S_1$-$S_2$ rates are not in strong tension with one another, however a proper comparison will require a detailed distorted wave calculation that is beyond the scope of this work. Unlike Scenario (a), the parameter space that is being probed by the $S_2$ only and $S_1$-$S_2$ analyses is hitherto unconstrained parameter space. Also shown in gray are terrestrial limits (a combination of constraints from Super-Kamiokande~\cite{Plestid:2020kdm}, LSND~\cite{Magill:2018tbb}, ArgoNeuT~\cite{Acciarri:2019jly}), recent results from the milliQan demonstrator 
\cite{Ball:2020dnx}, and the ceiling for virial DM in surface experiments \cite{Emken:2019tni}. 

While the tension with the $S_2$ only analysis is weak, we find that the parameter space capable of producing a 53 event excess \cite{Aprile:2020tmw} in the  $S_1$-$S_2$ analysis does not reproduce the lowest energy bin well as shown in \cref{excessstuff}.  This inability to reproduce the lowest energy bin's event rate persists for all choices of mDM mass $m_\chi$ and charge. However, a terrestrial build up of thermalized DM as considered in \cite{Pospelov:2020ktu} could produce monochromatic binding energy release, when binding to Xenon thus explaining this excess.

As we have already emphasized in \cref{fluxes}, the flux of RBDM is enhanced by four to five orders of magnitude via magnetic retention relative to what one would expect from conventional cosmic ray boosted dark matter  \cite{Bringmann:2018cvk,Ema:2018bih, Bondarenko:2019vrb,Cao:2020bwd,Dent:2019krz,Cappiello:2019qsw}. We demonstrate the consequences of this enhancements explicitly in \cref{fig:rbdmcomp-1} for $f_\chi=0.004$ and a hypothetical experimental sensitivity that is 1000 times as large as the current XENON1T data release with the same constraint of 5 events in this 0.5-1 keV bin. Compared to Scenario (a) we find that the effect of retention is much more dramatic in the $m_\chi-\epsilon$ plane. Naively the scaling with respect to $\epsilon$ should be identical between Scenario (a) and Scenario (b) since the production cross section scales as $\epsilon^2$ and both fluxes experience the same rigidity dependent magnetic retention factor,  however this turns out not to be the case. The reason for the unexpected scaling of signal with respect to $\epsilon$ stems from the IR-enhanced RBDM flux and effects related to the overburden which we discuss in \cref{attenuation}. As noted above, we emphasize that the "No Magnetic Retention" line can naturally arise in models with a light, but not massless, dark photon mediator whose mass shields the galactic magnetic fields, turning off retention. Detection in XENON1T would be phenomenologically identical to an mCP, but the resultant flux would be roughly five orders of magnitude smaller. 

This unexpectedly large sensitivity to the millicharge, $\epsilon$, means that XENON1T's current sensitivity, just beyond the limits from Super-Kamiokande and ArgoNeuT, can be pushed well into regions of unexplored parameter space with relatively modest improvements. To illustrate this point we  show limits for mCPs with fixed $f_\chi=0.004$ but with $f_\chi \times$(exp) increasing  \cref{fig:rbdmcomp-2} with exp standing for experimental sensitivity.  A large value of $f_\chi \times \textrm{exp}$ could occur either from a larger exposure (assuming fixed backgrounds), a large value of $\ell_\text{LB}$ (implying larger magnetic than assumed here), or if some other exotic physics were to source a fast flux of mCPs.  The advantageous scaling with respect to $\epsilon$, with signal scaling as $\epsilon^{1.86}$ [\emph{c.f.} \cref{scaling}], means that even modest improvements in experimental sensitivity, and/or modelling of magnetic retention will allow XENON1T to probe large swathes of as-of-yet unexplored parameter space in the near future. 

\begin{figure}
    \centering
    \includegraphics[width=\linewidth]{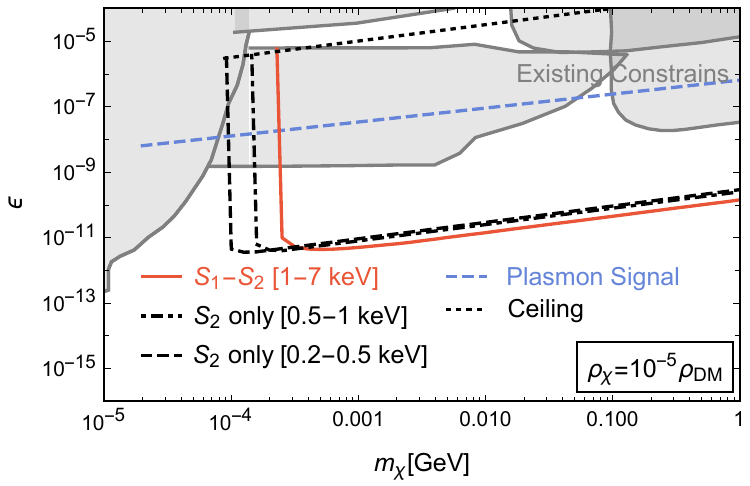}
    \caption{Sensitivity of XENON1T to a monoenergetic flux of mCPs moving with a velocity of $\beta_\chi = 0.1$. We find that XENON1T robustly excludes an mCP-induced plasmon-excess (as proposed in \cite{Kurinsky:2020dpb}) because $\beta_\chi=0.1$ is sufficiently fast to penetrate the overburden at Gran Sasso and kick electrons above the XENON1T thresholds. For lower velocities $\beta\leq 0.02$, the electron recoil drops below the thresholds for XENON1T's $S_2$-only analysis. Since a plasmon excitation requires $\beta_\chi \gtrsim 0.00675$, it is possible for a fast flux of mCPs to explain the plasmon-excess claimed in \cite{Kurinsky:2020dpb} without being in tension with XENON1T provided $0.00675\leq \beta_\chi \leq 0.02$.
    \label{fast-mono-flux}}
\end{figure}

\subsection{XENON1T and a mono-energetic flux of mCPs\label{mono-energetic}}
Although physically unrealistic, a monochromatic flux of mCPs is a useful example to work through because it fixes the velocity distribution and so is less sensitive to overburden attenuation. Motivated by the recent proposal of a fast sub-component of mDM as an explanation of low-energy excesses at a variety of experiments we study a benchmark velocity of $\beta_\chi \gamma_\chi = 0.1$. We update bounds from XENON1T (those presented in Fig.\ 6 of \cite{Kurinsky:2020dpb} assume a virial velocity and rely on nuclear recoils), and include the ceiling from overburden attenuation for XENON1T (3650 mwe).

Our results show that any parameter space capable of explaining the excess events observed at e.g.\ EDELWEISS via energy deposition into a collective plasmon mode, would necessarily overpopulate the $S_2$-only and $S_1$-$S_2$ bins from XENON1T.  This statement is dependent on $\beta_\chi$, and for small enough values of $\beta_\chi$ need not hold true. The electron recoil threshold at XENON1T naively sets a minimum velocity for the incident mCP of $\beta_\chi \gtrsim 0.02$. Excitation of a bulk plasmon does not require quite as fast an mCP with a velocity threshold of roughly $\beta_\chi \gtrsim 0.007$ \cite{Kurinsky:2020dpb}.  Therefore, for an electron energy loss spectroscopy (EELS) like mechanism to be at the source of the low-energy excess in EDELWEISS, the incident mCP must have a velocity satisfying $0.007 \lesssim \beta_\chi \lesssim 0.02$. This neglects the possibility of a coherent atomic recoil which requires a more sophisticated treatment of the atomic wavefunctions.  Searches for these events could in principle further tighten the window of acceptable velocities, but would require a more sophisticated treatment that is beyond the scope of this work.

\section{Overburden attenuation \label{attenuation} } 

Since XENON1T has low recoil thresholds, it is possible for non-relativsitic mCPs to leave detectable signatures. The slowest possible velocity in the $S_2$-only analysis \cite{Aprile:2019xxb} (with $T_e\sim 200$ eV) is $\beta_\chi \gamma_\chi \sim 0.02$. At such slow velocities, however, the overburden can stop mCPs for $\epsilon^2/ m_\chi \gtrsim 10^{-12}$ MeV$^{-1}$. This is especially important in the context of RBDM whose flux is dramatically enhanced as $\beta_\chi\gamma_\chi\rightarrow 0$.

To account for this effect we make use of range tables for protons passing through water \cite{pstar-range-tables}, and use XENON1T's overburden thickness quoted in meters water equivalent. Since $\langle \dd E / \dd x \rangle$ is a function of $\beta\gamma$ alone, we re-express this quantity as 
\begin{equation}
    \left\langle \dv{\beta\gamma}{x} \right\rangle =  \left\langle \dv{E}{x} \right\rangle \times \dv{\beta\gamma}{E} = -\frac{\epsilon^2}{m_\chi} g(\beta\gamma). 
\end{equation}
We then define an ``effective range'' via the continuous slowing down approximation as 
\begin{equation}
    \widetilde{\Delta x}(\beta\gamma_i) = \int_{\beta\gamma_\text{thr}}^{\beta\gamma_i} \frac{1}{g(\beta\gamma)} \dd \beta\gamma ~,
\end{equation}
where $\beta\gamma_\text{thr}$ is the threshold value of the experiment under consideration (for instance at XENON1T $\beta\gamma_\text{thr}$ ranges from 0.02 for the $S_2$-only analysis to 0.1 for the $S_1$-$S_2$ analysis). The true range, $\Delta x$ (in water), is then given by 
\begin{equation}
    \Delta x(\beta\gamma_i)= \frac{\epsilon^2}{m_\chi} \widetilde{\Delta x}(\beta\gamma_i)~. 
\end{equation}
Inverting this expression gives a minimum initial values of $\beta_\chi \gamma_\chi$ as a function of $\epsilon^2/m_\chi$, and we cut-off all of our spectra below this value. Our calculation agrees well with the similar $\beta \gamma_{\rm min}$ analysis in \cite{Dunsky:2018mqs}, as well as the overburden attenuation for the specific case of virial DM ($\beta \gamma \approx 10^{-3}$) in \cite{Emken:2019tni}. 

To a very good approximation the overburden serves to cut-off the flux at $\beta\gamma_\text{min}$ (we have checked by explicitly migrating the flux using the continuous slowing down approximation). Naively this truncation reduces the rate by an $\epsilon$ independent pre-factor, however this intuition fails for IR-peaked differential fluxes like the one predicted by RBDM. Instead the overburden attenuation changes the scaling of the signal with respect to $\epsilon$. 

\begin{figure}
    \includegraphics[width=\linewidth]{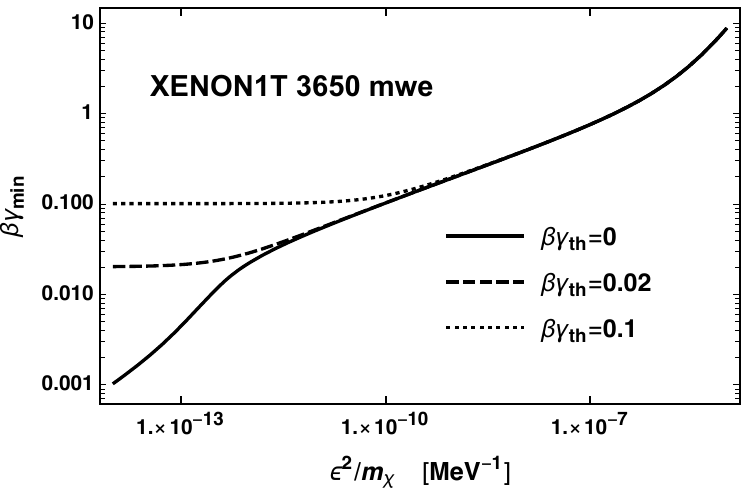}
    \caption{Minimum required $\beta\gamma$ for an mCP to penetrate 3650 mwe (XENON1T) of overburden as a function of $\epsilon^2/m_\chi$ for different values of initial $\beta\gamma_\text{th}$. Calculated using PSTAR proton range tables \cite{pstar-range-tables} in water and the continuous slowing down approximation.  } 
    \label{bgmin}
\end{figure}

The naive signal scaling is $\epsilon^2 \times \epsilon^{1/3} \times \epsilon^2 $ with factors coming from the production cross section, magnetic retention, and detection cross sections respectively. Notice, however, that $\beta\gamma_\text{min}$ is a function of $\epsilon^2/m_\chi$, and that the event rate scales as $\int_{\beta_\text{min}} \dd \beta~ \mathcal{I}_\chi^\text{RB}(\beta)/ \beta^2$ where the $1/\beta^2$ comes from the non-relativistic limit of Coulomb scattering.  Since the RBDM intensity scales  as $1/\beta^{3.33}$ this gives a signal that scales as  $(1/\beta\gamma_\text{min})^{4.33}$.  In the region where $10^{-11}\gtrsim\epsilon^2/m_\chi \lesssim 10^{-7}$ we find that $\beta\gamma_\text{min} \sim (\epsilon^2/m_\chi)^{0.285}$, such that the signal scales as 
\begin{equation}
\begin{split}\label{scaling}
    \text{signal} &\sim \epsilon^2 \times \epsilon^{1/3} \times \epsilon^2 \times (1/\beta\gamma_\text{min})^{4.33} \\
    &\sim 
    \epsilon^{4+1/3} \epsilon^{-2 \times 0.285 \times 4.33}\\ &\sim \epsilon^{1.86} \quad,\quad\qq{for Scenario (b)~.}
\end{split}
\end{equation}
This scaling relation of signal with respect to $\epsilon$ eventually breaks as $\beta\gamma_\text{min}$ asymptotes to $\beta\gamma_\text{thr}$ as shown in \cref{bgmin}. 

\begin{figure}
    \centering
    \includegraphics[width=\linewidth]{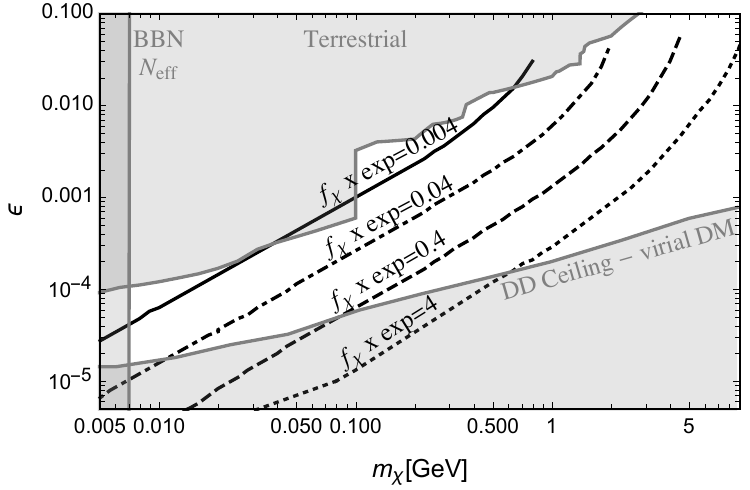}
    \caption{Scenario (b). Projections are shown for different $f_\chi \times \textrm{exp}$ with exp standing for experimental sensitivity.  Increased experimental sensitivity could be achieved either via detector improvements (e.g.\ longer exposure) or because of an enhanced flux of RBDM relative to our conservative estimate. For example a local over-concentration of mDM in the galactic disk can enhance the flux by a factor of 20 relative to our estimates (see \cref{different-scenarios}). Limits correspond to XENON1T $S_2$ only $0.5-1$ keV bin.} 
     \label{fig:rbdmcomp-2}
\end{figure}

This unintuitive scaling of signal with respect to $\epsilon$ paints a highly optimistic future for the discovery (or exclusion) of RBDM using underground detectors. Rather than the rapidly falling sensitivity of $\epsilon^{4.33}$ that one would naively expect, the $\epsilon^{1.86}$ scaling derived above is extremely advantageous. Decreases in overburden have a non-negligible impact on signal strength. We estimate that 300 mwe of overburden would increase the signal by a factor of 50-100 relative to the 3650 mwe at Gran Sasso, however it is unclear if this is a net gain in sensitivity given the increased cosmogenic activity in the detector. Even without adjusting overburden, we expect that XENON1T can significantly extend its sensitivity. Larger iterations of large noble gas detectors (i.e.\ future XENON-n-ton experiments) will have the advantage of both increasing exposure, and benefiting from enhanced self-shielding. Conventional mDM searches assume a virial velocity distribution, and their sensitivity is therefore limited in many cases by an overburden ceiling at at $\epsilon\sim 10^{-4}$. In contrast RBDM is sufficiently fast to both penetrate the overburden at Gran Sasso and to overcome XENON1T's thresholds, such that there may be room for XENON1T to probe significant regions of the as-of-yet unexplored parameter space shown in \cref{fig:rbdmcomp-2}. 

\section{Conclusions \label{conclusions}}

The advent of sub-keV direct detection thresholds in large scale detectors (like XENON1T) brings qualitatively new physics to the forefront of mCP direct detection. Of fundamental importance is the realization that for $m_\chi \gg m_e$,  the electron recoil threshold sets a limit on the \emph{velocity} of mCPs. For recoil thresholds in the sub-keV regime, mCPs can be non-relativistic and this has three important consequences. First, the detector response is sensitive to the atomic physics of the target material and care must be taken to properly address this issue (see \cref{atomic}). Second, the confluence of the IR-enhanced Rutherford cross section $\sigma \sim \alpha^2/\beta^2$ and an IR-dominated flux can lead to rates that are incredibly sensitive to the threshold energy (much more so than the naive $1/E_\text{thr}$ based on the Rutherford cross section would suggest). Finally, this sensitivity to the low-velocity components of an incident mCP flux can alter the parametric dependence of the rate with respect to $\epsilon$.  The overburden determines the minimum velocity of the mCP spectrum that ultimately reaches the detector. This minimum velocity is, in turn, determined by the charge-squared to mass ratio $\epsilon^2/m$ and therefore induces a non-trivial $\epsilon$ dependence in the event rate. In the case of RBDM studied here, this leads to a rate that scales as $\epsilon^{1.86}$ rather than the naive $\epsilon^{4.33}$ that one would expect after accounting for magnetic retention. 

This leads naturally to the second major conclusion of our study,  which is that diffusive motion of mCPs can dramatically increase the flux of mCPs arriving to Earth, to the point where the ISM can outshine the upper atmosphere. The mechanism of producing RBDM outlined here is robust, and is only mildly sensitive to various mDM transport hypotheses, and the RBDM subcomponent is best conceptualized as an incoherent gas, and is therefore insensitive to collective damping mechanisms; this differentiates RBDM from mDM that has been Fermi accelerated. Cosmic ray collisions with mDM will take place provided the mDM is located anywhere within roughly 15-20 kpc of the local disk, and the intensity of RBDM arriving at Earth only varies by a factor of a few for different mDM configurations. The largest achievable flux from redistributing mDM occurs when the mDM occupies the volume of the disk, and this results in an enhancement by a factor of roughly ten as shown in \cref{different-scenarios}. 

We find that XENON1T can realistically compete with large-scale neutrino detectors such as Super-Kamiokande in searches for mCPs produced via inelastic $pp$ collisions in both the upper atmosphere and the ISM. This minimal ``Scenario (a)'' offers a direct probe of mCPs independent of any dark matter assumptions. Our modelling of the magnetic transport of mDM is crude, and this subject deserves a more sophisticated treatment given that XENON1T and Super-Kamiokande could both benefit from a more precise prediction of the flux from the ISM (the flux from the upper atmosphere can already be reliably calculated).

On the mDM front, our study shows that a RBDM sub-component exists in any mDM model, and that this flux can be estimated, modeled, and its resultant sensitivity estimated. There are a number of uncertain ingredients in our formalism, however our estimates are relatively conservative: we make use of empirical measurements of the cosmic ray spectrum at low energies, we use a leaky box size of 10 kpc (rather than the 20-200 kpc suggested by \cref{different-scenarios}), and we have included reasonable treatments of the overburden attenuation and low-energy atomic physics governing the detector response. We find that XENON1T specifically is capable of probing RBDM, and that with increased exposure it can likely probe large swaths of parameter space out of reach of conventional direct detection experiments. This sensitivity reaches down ``from above'', i.e. for $\epsilon\geq$(the direct detection ceiling) in contrast to conventional direct detection which is limited by an overburden ceiling; searches for RBDM may then close the gap between accelerator and fixed target searches for mCPs and virialized mDM searches in underground experiments. The scaling of signal with respect to $\epsilon$ is especially advantageous because of the IR-enhanced nature of the RBDM flux. As one probes smaller values of $\epsilon$, larger fractions of the slow (and consequently large) RBDM flux are able to penetrate the overburden.

In conclusion, we have shown that mCP production in the ISM, magnetic retention, low detection thresholds, non-relativistic enhancements of Rutherford scattering, and non-trivial atomic physics all play a role in determining the sensitivity of the XENON1T experiment to mCPs. These conclusions apply especially to RBDM because it is the slowest available particles that dominate the rate. In this scenario overburden attenuation introduces a non-trivial modification of the signal scaling, and a proper account of the low-energy atomic physics is essential; this issue demands more attention. Our treatment of magnetic retention can be substantially refined, and we hope to inspire future work in the cosmic ray community in this direction. A more precise prediction of the RBDM spectrum would allow XENON1T unprecedented access to the lingering window of viable parameter space for mDM and push the experiment to the forefront in the search for mDM.   

\section{Acknowledgements} 
 We would like to thank Drs.\ B.\ Fields and I.\ Cholis for helpful discussions about cosmic ray transport. R.P.\ is extremely grateful for the  hospitality of the Fermilab theory group and acknowledges support from the Intensity Frontier Fellowship program which supported his visit to Fermilab. M.P.\ is grateful to Drs.\ T.\ Bringmann, J.\ Pradler and Y.-D.\ Tsai for useful discussions. R.P.\ was supported by the U.S. Department of Energy, Office of Science, Office of High Energy Physics, under Award Number DE-SC0019095. This manuscript has been authored by Fermi Research Alliance, LLC under Contract No.\ DE-AC02-07CH11359 with the U.S.\ Department of Energy, Office of Science, Office of High Energy Physics.  M.P.\ is supported in part by U.S. Department of Energy (Grant No.\ desc0011842).

\appendix 
 
 \section{Local millicharge intensity as a steady-state of diffusive transport \label{Diffusion-Discussion}}
 
We consider a phase space density of targets (either interstellar gas or mDM particles) $f_T(\bf{x},\bf{k})$ normalized such that 
\begin{equation}
   \int \dd^3{k} ~ f_\text{T}(\vb{x},\vb{k})  = n_\text{T}(\vb{x},\vb{k}) ~. 
\end{equation}
Similar definitions exist for the phase space density of cosmic rays, and kicked or produced mCPs. For a particle species $i$ with a non-zero velocity, the phase space density $f_i$ can be related to the local intensity (flux per steradian per unit momentum)
\begin{equation}
    \mathcal{I}_i(k,\Omega_k,\vb{x})= v(k)|\vb{k}|^2 f(\vb{k}, \vb{x})~,
\end{equation}
such that 
\begin{equation}
    \int \dd k \dd \Omega_k \mathcal{I}_i(k,\Omega_k,\vb{x}) = \langle n_i(\vb{x})v\rangle~.
\end{equation}
Then, the rate of production for mCPs per unit volume is 
\begin{equation}
   \dv{t} n_{\chi}= \int \dd^3{k}  \dd^3{p} f_T(\vb{x},\vb{k})  f_\text{CR}(\vb{x},\vb{p})  v_\text{rel} \sigma ~.
\end{equation}
If we consider the limit where $f_T(\vb{x},\vb{k}) = n_T(\vb{x})\times \delta^{(3)}(\vb{k})$ (i.e.\ treating all of the targets as at rest) and identifying $v_\text{rel} p^2 f_\text{CR}= \mathcal{I}_\text{CR}$ we then find
\begin{equation}
    \dv{t} n_\chi= n_T(\vb{x})\int_0^\infty   \dd p 
    [4\pi \mathcal{I}_\text{CR}(\vb{x},p) ] \sigma  ~.
\end{equation}
where we have made use of the fact that the cosmic ray intensity is isotropic. If we would like the intensity of mCPs instead, we find
\begin{equation}
     \dv{t} \mathcal{I}_{\chi}(k,\vb{x})= v(k) n_\text{DM}\int_0^\infty   \dd p 
        \mathcal{I}_\text{CR}(\vb{x},p)      
    \frac{\dd\sigma}{\dd k}  ~,
\end{equation}
where the factor of $4\pi$ has disappeared because the intensity is per-steradian. Having calculated the rate of production per unit volume, we now turn our attention to the resultant steady-state intensity distribution.

As a point of comparison, let us first consider uncharged DM (with a phase space density $f_0$), which undergoes convective (i.e.\ ballistic) transport $\dv{f_0}{t}= \partial_t f_0 + \vb{v} \cdot \nabla f_0$ after collisions with cosmic rays (see \cite{Bringmann:2018cvk,Bondarenko:2019vrb,Cao:2020bwd,Dent:2019krz,Cappiello:2019qsw,Dent:2020syp} for a some recent literature on this topic). Since $\mathcal{I}\propto f_0$ the intensity also satisfies a convective transport equation. The steady-state solution is defined via $\partial_t \mathcal{I}_0=0$, such that the steady-state boosted DM intensity is given by  
\begin{equation}\label{ballistic} 
    \vb{v}(k) \cdot \nabla \mathcal{I}_0(k,\vb{x}) =  v(k) n_\text{DM}\int_0^\infty   \dd p 
    \mathcal{I}_\text{CR}(\vb{x},p)
    \dv{\sigma}{k} ~.
\end{equation}
Then, integrating along the line of sight defined by $\hat{\vb{v}}$ we find 
\begin{equation}
    \mathcal{I}_0= \int_\text{LOS} \dd z  \int_0^\infty \dd p~ n_\text{DM}  \mathcal{I}_\text{CR}  \frac{\dd\sigma}{\dd k}~.
\end{equation}
If one takes $\mathcal{I}_\text{CR}$ to be independent of position then this can be written equivalently as
\begin{equation}
      \mathcal{I}_0(k,\vb{x}_\oplus) = \int_0^\infty \dd p~ \langle n^\perp_\text{DM}\rangle   \mathcal{I}_\text{CR} \frac{\dd\sigma}{\dd k }  ~, 
\end{equation}
in agreement with \cite{Bringmann:2018cvk}, where $\langle n_\text{DM}^\perp \rangle$ is the line-of-sight integrated column density.

The presence of strong magnetic fields in our galaxy means that mCPs move diffusively rather than ballistically\footnote{Only for $q/m \leq 10^{-12} ~\text{GeV}^{-1}$ do mCPs undergo ballistic rather than diffusive transport \cite{Stebbins:2019xjr}. } and therefore $\dd \mathcal{I}/\dd t = \partial_t \mathcal{I} + \tfrac12 D \nabla^2 \mathcal{I}$. Setting $\partial_t \mathcal{I}=0$ to find the steady-state solution, we arrive at a similar equation to \cref{ballistic}, but with the term on the left-hand side arising from diffusive transport
\begin{equation}\label{diffusive} 
   \frac{D(k)}{2} \nabla^2 \mathcal{I}_{\chi} = v(k) n_\text{T}\int_0^\infty   \dd p 
     \mathcal{I}_\text{CR}(\vb{x},p) 
    \dv{\sigma}{k} ~.
\end{equation}
We note here, that we are implicitly assuming a slow distribution of initial targets, however the generalization to a finite-speed distribution is obvious from the above discussion. If we impose boundary conditions at infinity in analogy with electrostatics then we can make use of the ``point-charge'' Green's function\footnote{For the Greens function appropriate for boundary conditions on a cylindrical boundary (as is often assumed in the disk diffusion model) see \S III of \cite{Ginzburg:1990sk} .}. Let us multiply through both side by $|\vb{v}|$ (the velocity of the RBDM) and $|\vb{k}'|^2$ such that the intensity of RBDM at the location of the earth is given by 
\begin{widetext}
\begin{equation}\begin{split}\label{generic-steady-state}
    \mathcal{I}_\chi(k',\vb{x}_\oplus) &= \frac{2 v(k') }{D} \int_0^\infty \dd p \dv{\sigma}{k'} \int \dd^3{y} 
    \frac{ \mathcal{I}_\text{CR}(\vb{y},p) n_T(\vb{y})}{4\pi |\vb{y}|} \qq{}\qq{with}\qq{}\vb{y}= \vb{x}-\vb{x}_\oplus\\
    &=\qty{\int_0^\infty \dd p \dv{\sigma}{k'}  \mathcal{I}_\text{voy}(p)} \times \qty{ \frac{2 v }{D}  \int \dd^3{y}~ \frac{\mathcal{N}_\text{CR}(\vb{y}) \times  n_T(\vb{y})}{4\pi |\vb{y}|}} ~, 
\end{split}\end{equation}
\end{widetext}
where in the second line we have assumed that $\mathcal{I}_\text{CR}(\vb{y},p)=\mathcal{N}_\text{CR}(\vb{y}) \mathcal{I}_\text{voy}(p)$ with the subscript ``voy'' referring to the spectrum measured by Voyager I  outside the heliosphere \cite{Cummings:2016pdr}. 

Notice that the second quantity in curly braces has the same units as a column density and can be interpreted as an ``effective column desnity'' that accounts for the enhancement of the density for a system undergoing diffusive transport relative to ballistic transport. 
\begin{equation}
    n^\perp_\text{eff}= \frac{2 c}{D_0}  \int \dd^3{y}~ \frac{\mathcal{N}_\text{CR}(\vb{y}) \times  n_T(\vb{y})}{4\pi |\vb{y}|}~.
\end{equation}
where we have implicitly assumed a magnetic rigidity of 1 GeV, and used
\begin{equation}
    D= D_0 \beta \qty(\frac{R}{1~\text{GeV}})^\delta ~,
\end{equation}
such that $v/D= c/D_0 \times R_\text{GeV}^{-1/3}$ (with $R_\text{GeV}$ a shorthand for $\frac{R}{1~\text{GeV}}$). The magnetic retention factor is defined as the ratio of the flux in the diffusive regime relative to the naive ballistic-transport estimate involving a line-of-sight column density
\begin{equation}
   \mathcal{R}_B=  \frac{n^\perp_\text{eff}}{\langle n_T^\perp \rangle_\text{LOS}} ~.
\end{equation}
Comparing to \cref{mag-ret-eq}, we can \emph{define} the leaky-box size, in a specific model of our target and cosmic-ray distribution as
\begin{align}
    3 \times 10^4 \qty[\frac{\ell_\text{LB} }{10~\text{kpc}}]&= \frac{2 c}{D_0} \frac{\int \dd^3y ~ (\mathcal{N}_\text{CR}\times  n_T)\big/(4\pi |\vb{y}|)}{\langle n_T^\perp \rangle_\text{LOS}}
    \nonumber 
\end{align}

If we consider the ISM production in Scenario (a), then we can express the target number density as  $n_T= \mathcal{N}_T \times (1$ cm$^{-3}$), and re-express the pre-factor in units of kpc such that 
\begin{equation}\label{LB-diff-def}
    \ell_\text{LB}=\frac{1}{0.5~\text{kpc}}\int \dd^3y ~ \frac{\mathcal{N}_\text{CR}(\vb{y})\times  \mathcal{N}_T(\vb{y})}{4\pi |\vb{y}|}~,
\end{equation}
where $\mathcal{N}_\text{CR}$ and $\mathcal{N}_T$ are dimensionless functions defined as the ratio of the local cosmic ray intensity to the voyager measurement, and the ratio of the local target density. Notice that, as defined in \cref{LB-diff-def}, $\ell_\text{LB}$ can depend on the target under consideration. For instance, if mDM has a very different spatial distribution than the interstellar gas, then a different choice of $\ell_\text{LB}$ will be required to describe the magnetic enhancement of the ISM and mDM components of the flux as illustrated explicitly in \cref{different-scenarios}.

As discussed in \cref{large-fluxes}, the flux of mDM arriving can be sourced by mechanisms other than Rutherford scattering, however the spatial distribution of mDM alone can also lead to an enhanced flux.  For example, if there are overlapping regions where both the CR intensity and the mDM density are simultaneously overdense, then this can lead to a larger $\ell_\text{LB}$ as defined in \cref{LB-diff-def}. Similarly, if the low-energy CR intensity is larger than what Voyager has measured (as was expected in some models), then this could also enhance the low-$\beta_\chi$ part of the RBDM flux (which dominates the event rate estimates at XENON1T).

To make things concrete we calculate the integral in  \cref{large-fluxes} for three different profiles of $\mathcal{N}_T$ in \cref{LB-diff-def} with $\mathcal{N}_\text{CR}$ held fixed. These choices are meant to be illustrative rather than realistic, and should only be interpreted as a useful exercise for gaining intuition. Nevertheless they give some sense of a credible interval of values that $\ell_\text{LB}$ can assume. 
\begin{enumerate}
    \item The mDM's number distribution follows a spherically NFW profile. 
    \item The mDM is coupled to baryons and follows the baryonic density exactly; all of mDM is located inside the disk.  
    \item The mDM has been evacuated into two disk-shaped pancakes each containing half the mDM population, with both pancakes displaced some $\sim 5$ kpc  perpendicular to the disk. 
\end{enumerate}
In each of these scenarios we can calculate the intensity of mDM arriving at Earth as predicted by \cref{generic-steady-state} for a rigidity of $R=1$ GeV, and then infer the equivalent leaky-box length used to parameterize the results in the main text.  

For the cosmic ray density profile we have produced a (very rough) interpolation of Fig.\ 11 of \cite{Cholis:2011un}. The explicit functional form is given by 
\begin{widetext}
\begin{equation}
    \mathcal{N}_\text{CR}(r,z)=14.2\left[\frac{0.2 \left(1-\e{-\left(\frac{r}{15}\right)^2}\right)}{r^{0.9}}+\e^{-\left(\frac{r}{6.8}\right)^2}\right] \e^{ -\frac{1}{4} \left[\left(\frac{z}{2.8}\right)^2+0.75 z+0.8\right]} \left[\tfrac{1}{8} r~ \e^{-\frac{z}{4}}+1\right]~,
\end{equation}
\end{widetext}
which has been written in cylindrical coordinates with $r$ and $z$ expressed in kpc. We take the Earth to be located at $r_\oplus=8$ kpc, and $z_\oplus=0$.  

We perform the same exercise for Scenario (a) using a disk-like exponential profile for the ISM density
\begin{equation}
    n_\text{ISM}(r,z)=(14.4~\text{cm}^{-3}) \exp\qty[-\frac{\abs{z}}{0.7~\text{kpc}} - \frac{r}{3~\text{kpc}}]~.
\end{equation}
The normalization is chosen such that the ISM density at Earth's location is 1 cm$^{-3}$. Using this profile, we estimate that the effective column density is $1.14 \times 10^{5}$ kpc/cm$^3$, corresponding to a leaky-boxy length of 19 kpc.

\section{Adjusting the free electron cross section to describe atomic ionization \label{atomic}}

\begin{table}
\centering
$\begin{array}{ ccc}
\text{Scenario.} 		& n^\perp_\text{eff}~[\text{kpc/cm}^3] & \text{Equivalent~} \ell_\text{LB} ~[\text{kpc}] \\
\midrule
\text{NFW profile}		& 1.0\times 10^6	& 17		\\
\text{Disk profile}		& 1.2\times 10^7    & 200	\\
\text{Sandwich profile} & 9.7\times 10^6	& 16 . \\
 \bottomrule
\end{array}$
\caption{ Effective column density for a mCP with a rigidity of 1 GeV for the three scenarios outlined in the text. The sandwich profile is composed of two disk profiles, each with half of the mDM, located 5 kpc away from the galactic disk (out of plane). The NFW profile is normalized such that the local mass density of mDM equals $f_\chi\times \rho_\text{DM}$. The disk profile assumes $n_\text{mDM}(r,z)\propto \exp[-\abs{z}/(0.7~\text{kpc} - r/(3~\text{kpc})]$, and is normalized such that the mass of the disk is equal to $f_\chi$ times the mass of the Milk Way's dark matter population.    \label{different-scenarios} }
\end{table}

In this Appendix we describe our approach to the atomic ionization by mCPs. The Born approximation 
for our problem remains valid at all times due to the smallness of $\epsilon$, or more precisely due to $\epsilon \alpha /\beta_\chi \ll 1$ condition. The most precise prescription for treating atomic ionization  must result from atomic theory calculations and/or experimental measurements using ordinary Standard Model projectiles. While the former is computationally 
challenging for the problem at hand, the latter often exists, albeit in domains of energy/momentum transfers not directly applicable to our case. 

When the energy transfer to the electron much exceeds its binding energy, the scattering is quasi-free. Nevertheless, one can consider approximate treatments of the atomic physics for the needs of direct detection problem (see e.g.\ \cite{Essig:2011nj}). These include {\em i)} the replacement of the final state electron wave function by plane waves $\exp(i\bf{kr})$, as well as {\em ii)} the prescription where matrix elements are calculated using final state plane waves, but the cross section is additionally augmented by the so-called Fermi (or Sommerfeld) factor: 
\begin{equation}
    S = \frac{2\pi \alpha m_e}{k }\frac{1}{1-\exp\left(\frac{-2\pi \alpha m_e}{k }\right)},
\end{equation}
where for the inner shells $\alpha \to Z_{\rm eff} \alpha$ substitution is used. The reliability of these approximate methods for the xenon atom is questionable. In addition, if the plane waves are used in the final state that are not orthogonal to the bound electron wave function, an {\em ad-hoc} substitution in the transition operator, $\exp(i\bf{qr})\to \exp(i\bf{qr})-1$, is often used to restore the correct qualitative behavior of the transitional matrix element in the small momentum transfer $\bf{q}$ limit. 
In order to gain insight into this problem, we conduct the comparison of these methods in a case where there exists closed analytic formulae that takes into account exact outgoing wave functions, i.e.\ the hydrogen atom. The full expression for the ionization rate can be found in e.g.\ the textbook \cite{Landau:1991wop}, section 148, problem 4. 

\begin{figure}
    \includegraphics[width=\linewidth]{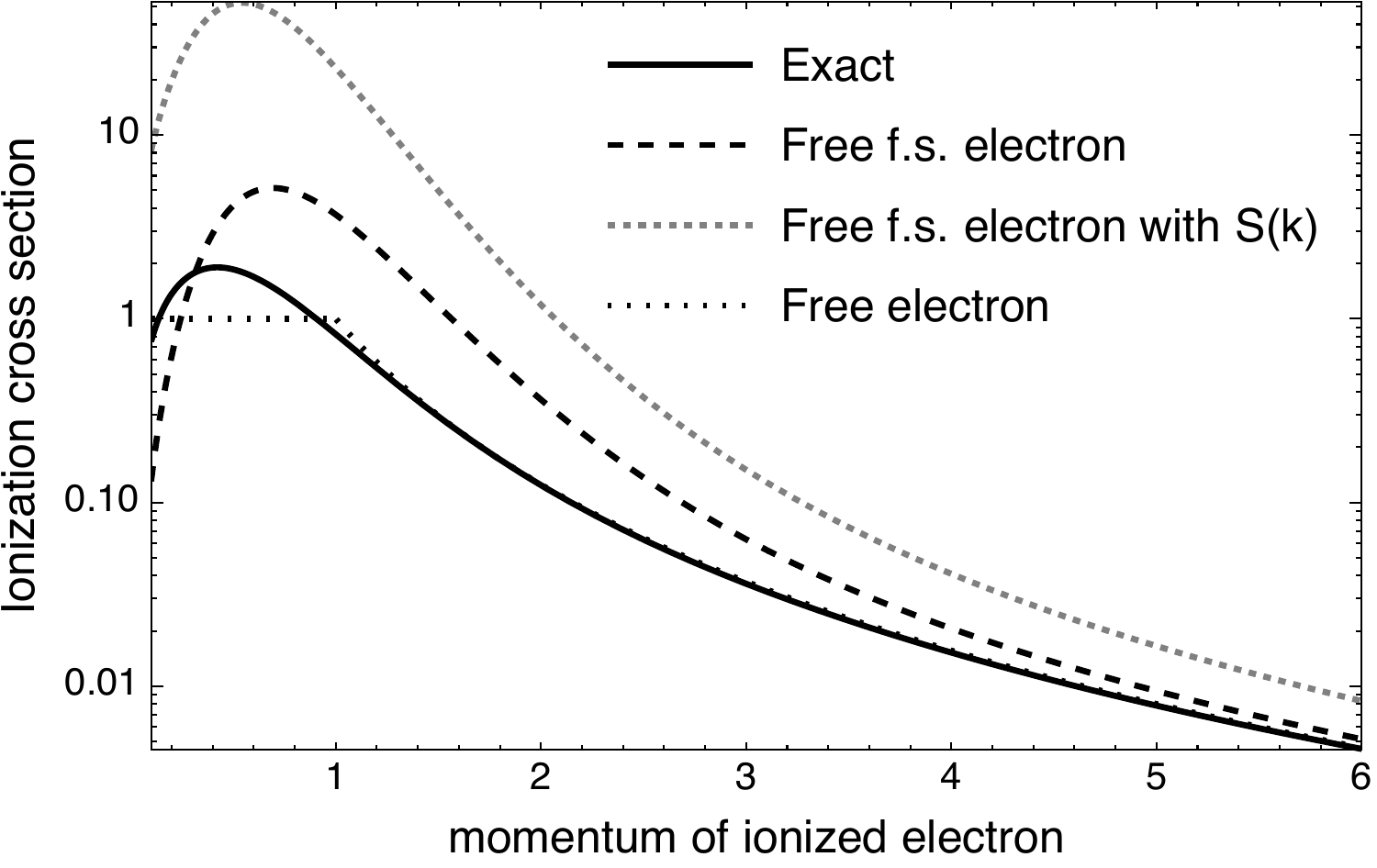}
    \caption{Ionization cross section of atomic hydrogen within various approximations. The Sommerfeld/Fermi factor [with $S(k)$] clearly overestimates the cross section at low momentum transfers. We use the piecewise approximation labeled ``Free electron''.  \label{cross-section}} 
   
\end{figure}

 \Cref{cross-section} shows the results for the hydrogen atom, as a function of the momentum of the ionization electron, using atomic units with $a_B = \frac{1}{\alpha m_e}=1;~m_e\alpha^2=1$, so that the energy of the final state electron is $k^2/2$ and the total energy absorbed by hydrogen is $(1+k^2)/2$. These cross sections, $d\sigma/dk$,  are integrated over all possible momentum transfers ${\bf q}$ and are given in units of $\sigma_0=8\pi  (m_e \alpha)^{-2} \left( \frac{\epsilon\alpha}{v_\chi} \right)^2$, where $v_\chi$ is the velocity of passing mCP particles. We show the full cross section (first order Born approximation with respect to the incoming particle, with exact accounting of the ionization electron final state wave functions),
as well as the results of the approximations {\em i} and {\em ii}. (In both approximations, we use the already-mentioned ``subtract 1'' prescription.) We see that in both cases the approximate cross section represents a significant overestimation in the $k\sim O(1)$ region. In particular, the prescription {\em ii} overestimates the peak cross section by over one order of magnitude. At the same time, we also plot the result for the cross section on free and static electrons that has $d\sigma/dk \propto 1/k^3 $ scaling, which gives a good fit everywhere for $k>1$. In order to remove what would be an unphysical in atomic setting $1/k^3$ infrared tail, we replace the free cross section with the constant value at $0<k<1$. The resulting curve deviates from the full answer by at most a factor of 2, which we deem to be acceptable on the scale of other uncertainties involved elsewhere in our calculations. 

Therefore, we resort to employing the approximate treatment of the ionization, and for that purpose modify the cross section of mCP scattering on free electrons using the following prescription. Consider the ionization of an electron with binding energy $E_b$, resulting in the energy transfer to the atom $E_t$. When the kinetic energy of the final state electron becomes 
equal or larger than the initial binding energy, we are using scattering on free electrons. For the kinetic energy of final state electron $0\leq E_t-E_b\leq E_b$ we approximate the cross sections by a constant, and for $E_t < E_b$ the cross section is assumed to be zero:
\begin{eqnarray}
    \frac{d\sigma}{dE_t }(E_t,E_b) = 
    \begin{cases} 
     0 &{\rm if}~ E_t<E_b\\
     2\pi \left( \frac{\epsilon\alpha}{\beta_\chi} \right)^2    \frac{1}{m_e(E_b)^2} & {\rm if}~ E_b<E_t<2E_b\\
     2\pi \left( \frac{\epsilon\alpha}{\beta_\chi} \right)^2 \frac{1}{m_e(E_t-E_b)^2} & {\rm if}~ 2E_b<E_t 
    \end{cases}.
    \nonumber\\
    \label{prescr}
\end{eqnarray}
These formulae are written in the limit of $m_\chi \gg m_e$ and $m_\chi \beta_\chi^2/2 \gg E_t$. 
The total ionization cross section at a given energy transfer $E_t$ is obtained by summing over all shells:
\begin{equation}
    \frac{d\sigma}{dE_t }(E_t) = \sum_{b} N_b \frac{d\sigma}{dE_t }(E_t,E_b),
\end{equation}
where $N_b = 2J_b+1$ is a given shell electron multiplicity. The values of energies for xenon shells are taken from Ref.\ \cite{Dzuba:2010cw}. When $m_\chi \beta_\chi^2/2 \gg E_t$ is not assumed, we need to include an additional multiplicative factor $\left( 1-\beta_\chi^2 \frac{E_t-E_b}{E_{max}} +\frac{(E_t-E_b)^2}{2E_\chi^2}\right)$ in the third line of \cref{prescr}, and $\left( 1-\beta_\chi^2 \frac{E_b}{E_{max}} +\frac{(E_b)^2}{2E_\chi^2}\right)$ in the second line for continuity, where $E_{max}$ is the maximum energy that kinematically can be passed to a static electron by a passing particle of mass $m_\chi$ and velocity $\beta_\chi$. 

Given the log-log nature of the mCP plots explored in this paper, we believe that the accuracy provided by the above treatment is sufficient. If, however, an mCP flux, or any other light DM flux, is entertained to be behind recent XENON1T excess, a far more thorough atomic physics treatment is needed (possibly with multi-electron correlations taken into account), as the exact status of the excess vs exclusion limits may indeed be  quite sensitive to atomic details. 

\bibliography{biblio.bib}
\end{document}

%% file: Commands/ryansCommands.tex
\def\e{\mathrm{e}}
\def\nPerpISM{n_\text{ISM}^\perp}